\documentclass[11pt]{article}
\usepackage[letterpaper,margin=1in]{geometry}
\usepackage{setspace}
\usepackage[T1]{fontenc}
\usepackage{graphicx}
\usepackage{natbib}
\usepackage[english]{babel}
\usepackage{arydshln}
\usepackage{amsmath}
\usepackage{amsthm}
\usepackage{amssymb, soul}
\usepackage{mathtools}
\usepackage{bm}
\usepackage{enumitem}
\usepackage{tikz}
\usetikzlibrary{arrows.meta, positioning, fit, backgrounds, calc}
\usepackage{setspace}
\usepackage{threeparttable}
\usepackage{tabularx}
\usepackage{booktabs}
\usepackage{longtable}
\usepackage{arydshln}
\usepackage{multirow}
\usepackage{array}
\usepackage{mathabx}
\usepackage{appendix}
\usepackage{graphicx}
\newtheorem{theorem}{Theorem}
\newtheorem{lemma}[theorem]{Lemma}

\newtheorem{corollary}[theorem]{Corollary}
\usepackage[colorlinks=true,linkcolor=red,citecolor=blue,urlcolor=blue]{hyperref}
\usepackage[nameinlink,noabbrev]{cleveref}

\newcommand{\E}{\mathbb{E}}

\newcommand{\Var}{\operatorname{Var}}

\newcommand{\op}{o_p}
\newcommand{\Op}{O_p}

\providecommand{\keywords}[1]{\vspace{0.5em}\noindent\textbf{Keywords:} #1}

\title{Covariate Adjustment for Time-Restricted Marginal Win Statistics in Randomized Trials: Assumption-Lean Inference and Guaranteed Efficiency Gain}

\author{
  Baoshan Zhang\,$^{1}$
  \and Laine E. Thomas\,$^{1}$
  \and Yuan Wu\,$^{1}$
  \and Yu Du\,$^{2}$
  \and Tuo Wang\thanks{Corresponding author: \texttt{tuo.wang@lilly.com}}\,\,$^{,2}$
}
\date{\today}

\begin{document}
\maketitle

\noindent
$^{1}$\,Department of Biostatistics and Bioinformatics, Duke University, Durham, NC\\
$^{2}$\,Department of Global Statistical Sciences and Advanced Analytics, Eli Lilly and Company, Indianapolis, IN

\begin{abstract}
Hierarchical composite endpoints, analyzed through win statistics such as the net benefit (NB), win ratio (WR), and win odds (WO), are increasingly used in randomized trials to prioritize endpoints according to clinical importance. Although covariate-adjustment methods for win statistics have been developed, a unified model-assisted approach for mixed-type, time-restricted HCEs with simultaneous asymptotic no-loss efficiency guarantees remains of interest. We propose a unified, model-assisted covariate-adjustment framework for time-restricted marginal win statistics evaluated at a prespecified post-randomization restriction horizon, a clinically interpretable estimand particularly suited to acute and high-acuity trials with well-defined short-term follow-up windows. The Hoeffding--H\'ajek projection of the two-sample win--loss $U$-statistics yields subject-level influence functions for the marginal win and loss probabilities. We apply treatment-specific linear covariate adjustment to these influence functions and propagate the adjustment to NB, log WR, and log WO. The resulting estimators are \emph{assumption-lean}, remaining consistent and asymptotically valid without correct specification of the linear working model, and are asymptotically \emph{guaranteed efficiency gain} over their unadjusted counterparts. Simulations with mixed-type HCEs demonstrate substantial efficiency and power gains with prognostic covariate adjustment. However, including additional non-prognostic covariates can impair finite-sample variance estimation and inflate Type I error, particularly in smaller trials. Bootstrap variance estimation improved inferential calibration in these settings. We illustrate the proposed method using a randomized acute kidney injury alert trial.
\end{abstract}

\newpage

\section{Introduction}
Composite endpoints are widely used in randomized clinical trials to assess several clinically important endpoints simultaneously. Conventional time-to-first-event analyses 
consider only the first event and generally ignore differences in clinical importance among components. Thus, an earlier but less consequential event may determine the analysis while a later, more serious event such as death is disregarded \citep{ freemantle2003composite, redfors2020win}.  Hierarchical composite endpoints (HCEs) instead pre-specify the component priorities and comparing patients sequentially down the hierarchy, so that a lower-priority component is considered only when higher-priority components do not distinguish the pair. 
\cite{pocock2012win} proposed the Win Ratio (WR), part of a broader family of win statistics, including the Net Benefit (NB), and the Win Odds (WO), which share a pairwise win--loss--tie construction but differ in how they summarize the resulting win, loss, and tie probabilities \citep{dong_win_2023}. 

A central issue for win statistics is the choice of estimand. Two-sample win statistics target a marginal treatment contrast between groups, whereas model-based formulations focus on covariate-specific, conditional effects. \citet{mao2021class} introduced proportional win-fractions models for conditional WR regression and \cite{hazewinkel2026covariate} proposed an ordinal-logistic approach that likewise targets a conditional WR. These approaches estimate a conditional treatment effect but do not directly deliver the marginal effect that is typically the primary estimand of randomized clinical trials \citep{fda2023covariate}. 
Win statistics are generally non-collapsible, so conditional and marginal effects need not coincide \citep{dong2026win}. Moreover, the marginal estimand depends on the duration of follow-up and the pattern of censoring for time-to-event components \citep{oakes2016win, dong2020win, wang2025restricted}. 
Therefore, defining a marginal win estimand requires explicit specification of the follow-up time frame. In this project, we target a time-restricted marginal win estimand defined over a prespecified post-randomization horizon $S$, following the framework of \citet{mao2024defining}.

Although WR was first used mainly in cardiovascular trials, recent applications have extended to perioperative and critical-care settings \citep{fandino2025efficient}. A time-restricted formulation is especially natural in these settings, where treatment effects are often defined over short, clinically meaningful post-randomization windows. Fixing the horizon provides a well-defined and interpretable estimand rather than one induced by trial-specific follow-up or censoring \citep{mao2024defining}. For example, DEFENDER and ANDROMEDA-SHOCK-2 used WR to analyze hierarchical primary endpoints over 28-day horizons after randomization \citep{tavares2024dapagliflozin, andromeda2025personalized}, while two large acute kidney injury (AKI) trials evaluated AKI progression, dialysis, or death as primary outcomes within 14 days \citep{wilson2021electronic, wilson2023randomized}. These horizons reflect the expected course of acute illness rather than statistical convenience alone. Accordingly, we consider a prespecified clinical horizon $S$ and focus on settings with complete ascertainment of the HCEs through $S$, a plausible scope for short-horizon acute-care trials in which follow-up is intensive and loss to follow-up is uncommon, as in DEFENDER and ANDROMEDA-SHOCK-2.

Having fixed the time-restricted marginal estimand, we consider covariate adjustment to improve precision while preserving target estimand via prognostic baseline covariates, as encouraged by regulatory guidance \citep{fda2023covariate}. Here, we seek two complementary properties. First, adjustment should be \emph{assumption-lean}: consistency and asymptotically valid inference should not depend on correct specification of a working model \citep{bannick2026covariate, zhang2025unified}. Second, it should satisfy an asymptotic \emph{guaranteed efficiency gain}, wherein adjustment cannot increase variance relative to the unadjusted estimator, even when the working model is mis-specified \citep{ye2023toward}. This property is exemplified by ANHECOVA for mean of continuous outcomes, whose treatment-specific linear adjustment is asymptotic \emph{guaranteed efficiency gain} verse ANOVA under randomization 
\citep{lin2013agnostic, ye2023toward}. Our goal is to extend these two properties to marginal win statistics.

Several recent methods have extended covariate adjustment to win measures. 
\citet{wang2025adjusted} proposed inverse-probability-of-treatment-weighted WR estimators to address baseline covariate imbalance, while \citet{gasparyan2021adjusted} developed stratified and covariate-adjusted WR methods allowing adjustment for a single numerical covariate. More recently, \citet{cao2025covariate} developed inverse probability weighting and overlap weighting estimators for WR and NB with ordinal outcomes, establishing \emph{guaranteed efficiency gain} variance relative to unadjusted estimation. They also proposed augmented weighting estimators that remain consistent under outcome-model misspecification, although their guaranteed asymptotic efficiency gain requires a correctly specified outcome model. Moreover,  \citet{scheidegger2026covariate}, building on \citet{vermeulen2015increasing}, developed an assumption-lean covariate adjusted estimator of the marginal WO. However, their approach does not establish a \emph{guaranteed efficiency gain} over the unadjusted estimator under working model mis-specification.  Moreover, the WO is determined by the difference between win and loss probabilities (NB), whereas the difference alone does not determine the WR when ties are present. Thus, a unified framework for time-restricted WR, NB, and WO that accommodates mixed-type hierarchical outcomes and provides simultaneous asymptotic \emph{guaranteed efficiency gain} remains lacking.

In this work, we address this gap by exploiting the subject-level Hoeffding--H\'ajek projection of the two-sample win--loss U-statistics. The projection yields asymptotically linear subject-level contributions, to which treatment-specific linear covariate adjustment can be applied.  We jointly project the elementary win and loss influence functions onto a common linear space of baseline covariates and propagate the resulting adjustment to each win measure. By construction, it provides simultaneous asymptotic \emph{guaranteed efficiency gain} relative to unadjusted estimation for all win measures, even when the linear working models are mis-specified.  
Moreover, the framework accommodates protocol-defined mixed-type HCEs and provides closed-form large-sample inference. Our simulation study examines how different component-specific prognostic covariate-selection strategies translate into efficiency gains for marginal win estimands. We also investigate the finite-sample impact of including non-prognostic covariates and assess the calibration of analytical and bootstrap variance estimators.

The remainder of the paper is organized as follows. Section \ref{sec:unadjusted} defines the fixed-horizon hierarchical comparison, marginal win estimands, and unadjusted estimators. Section \ref{sec:Adjusted} develops the covariate-adjusted estimators and their asymptotic properties. Sections \ref{sec:simulation} and \ref{sec:realana} present the simulation study and real-data application, respectively.  Section \ref{sec:discussion} discusses limitations and future extensions.

\section{Time-Restricted Win Marginal Estimands and Unadjusted Estimators}
\label{sec:unadjusted}

\subsection{Data structure and fixed-horizon endpoint records}
\label{subsec:data}
Assume we observe $N$ independent and identically distributed (i.i.d.) subjects, $\mathcal O_i(S)=\{A_i,\mathbf X_i,\mathcal D_i(S)\}\sim\mathcal P,$ for $i=1,\ldots,N, $ where $A_i\in\{0,1\}$ is the treatment indicator ($A_i=1$ for treatment and $A_i=0$ for control), $\mathbf X_i$ is a vector of baseline covariates, and $\mathcal D_i(S)$ denotes the HCE record evaluated through a prespecified post-randomization restriction horizon $S$. The i.i.d.\ assumption is across subjects and imposes no restriction on within-subject dependence. Under $\mathcal P$, treatment is randomized independently of baseline covariates, with $A_i\perp\mathbf X_i$ and $\Pr(A_i=1)=e\in(0,1)$, while the distribution of $\mathcal D_i(S)$ may depend on both $A_i$ and $\mathbf X_i$. Let $n_1=\sum_{i=1}^N A_i$ and $n_0=N-n_1$ denote the treatment-specific sample sizes; by the law of large numbers, $n_1/N\to e$ and $n_0/N\to1-e$ in probability.

The restriction horizon \(S\) is the same elapsed follow-up time from randomization for all participants, rather than a common calendar study-end date. The HCE record ${D}_i(S)$ consists of $Q$ component endpoints, observed through time horizon $S$, ordered by clinical importance or severity, $\mathcal{D}_i(S) = \{ D_{1i}(S), D_{2i}(S), \ldots, D_{Qi}(S)\}.$   Components may differ in type, including binary, continuous, ordinal, count, duration, and time-to-event (TTE) outcomes. For a non-TTE component, $D_{qi}(S)$ denotes the corresponding endpoint summary defined through $S$, such as a binary event indicator by \(S\), a continuous or ordinal measurement at \(S\), or a count or duration accumulated through \(S\).  For a TTE component, let $T_{qi}$ denote the latent event time and define horizon-restricted event time $ \widetilde{T}_{qi}(S) = \min(T_{qi}, S)$ together with the event indicator $\delta_{qi}(S) = I(T_{qi} \le S)$. The corresponding fixed-horizon component record is $ \mathcal D_{qi}(S) =\{\widetilde T_{qi}(S),\delta_{qi}(S)\}.$ Thus, TTE comparisons use only information accrued through the common restriction horizon \(S\). We assume that HCE components are fully observed through \(S\), with no censoring or loss-to-follow-up before \(S\).  Such complete ascertainment is reasonable in settings with short $S$, protocol-defined horizons, as often encountered in acute and high-acuity trials. 

For notational simplicity, since the restriction horizon \(S\) is fixed throughout, we suppress explicit dependence on \(S\) hereafter when no ambiguity arises, writing \(\mathcal D_i=\mathcal D_i(S)\), \(D_{qi}=D_{qi}(S)\), \(\widetilde T_{qi}=\widetilde T_{qi}(S)\), and \(\delta_{qi}=\delta_{qi}(S)\). 

\subsection{Pairwise comparison rules and win--loss kernels}
\label{subsec:kernels}
Next, we define the pairwise comparisons between a treatment subject $i$ ($A_i = 1$) and a control subject $j$ ($A_j = 0$). Starting from the highest-priority component $q=1$, the pair is compared until one subject wins; if the component is tied, the comparison proceeds to the next component. If all $Q$ components are tied, the pair is declared an overall tie. For each component $q$, let $W_q^{ij}$, $L_q^{ij}$, and
$\Omega_q^{ij}$ denote the component-specific win, loss, and tie
indicators for treatment subject $i$ relative to control subject $j$,
with $ W_q^{ij}+L_q^{ij}+\Omega_q^{ij}=1.$ Table~\ref{tab:comparison_rules} summarizes common comparison rules for the endpoint types considered here. 
Under the default orientation in Table~\ref{tab:comparison_rules}, the absence of a binary adverse event, fewer adverse-event counts, shorter durations such as ICU stay, and lower severity scores are preferable, whereas for time-to-event adverse outcomes, later event times are preferable.

\begin{table}[ht]
\begin{threeparttable}
\caption{Common pairwise comparison rules for component-specific win, loss, and tie indicators.}
\label{tab:comparison_rules}
\footnotesize
\begin{center}
\begin{tabular}{llll}
\toprule
\textbf{Component data type} & \textbf{Win, $W_q^{ij}$} & \textbf{Loss, $L_q^{ij}$} & \textbf{Tie, $\Omega_q^{ij}$} \\
\midrule
Binary & $I(D_{qi}<D_{qj})$ & $I(D_{qi}>D_{qj})$ & $I(D_{qi}=D_{qj})$ \\ 
\addlinespace 
Continuous/Ordinal/Count/Duration & $I(D_{qi}<D_{qj}-\epsilon_q)$ & $I(D_{qi}>D_{qj}+\epsilon_q)$ & $I(|D_{qi}-D_{qj}|\le\epsilon_q)$ \\
\addlinespace
Time-to-event & $\delta_{qj}\, I(\widetilde{T}_{qi} > \widetilde{T}_{qj} + \epsilon_q)$ & $\delta_{qi}\,I(\widetilde{T}_{qi} < \widetilde{T}_{qj} - \epsilon_q)$ & $(1 - W_q^{ij})(1 - L_q^{ij})$ \\
\bottomrule
\end{tabular}
    \begin{tablenotes}\footnotesize
        \item For scalar non-TTE components, the table is written under the convention that larger values represent less favorable outcomes; inequalities are reversed when larger values are clinically preferable. For TTE components, later event times are assumed preferable. The constant $\epsilon_q\ge0$ is a prespecified relevance margin. Dependence on the fixed restriction horizon $S$ is suppressed as described in Section~\ref{subsec:data}.
    \end{tablenotes}
\end{center}
\end{threeparttable}
\end{table}

Aggregating the component-specific comparisons according to the prespecified hierarchy, the win and loss kernels are
\begin{equation}
\varphi_w(\mathcal{D}_i, \mathcal{D}_j) = W_1^{ij} + \sum_{q=2}^{Q}
\Big( \prod_{p=1}^{q-1} \Omega_p^{ij} \Big) W_q^{ij},
\qquad
\varphi_l(\mathcal{D}_i, \mathcal{D}_j) = L_1^{ij} + \sum_{q=2}^{Q}
\Big( \prod_{p=1}^{q-1} \Omega_p^{ij} \Big) L_q^{ij}.
\label{eq:kernel_functions}
\end{equation}

\subsection{Marginal Win Estimands, Unadjusted Estimators, and Influence Functions}\label{subsec:estimands}
All quantities in this subsection and the reminder of the paper are evaluated at the prespecified restriction horizon \(S\), whose dependence is suppressed for notational simplicity.  Denote \(\mathcal D_T\) and \(\mathcal D_C\) as independent HCE records drawn from treatment ($A=1$) and control group ($A=0$), respectively. The marginal win and loss probabilities are
\[
\tau_\nu
=
\E[\varphi_\nu(\mathcal D_T,\mathcal D_C)],
\qquad \nu\in\{w,l\},
\]
and overall tie probability $\tau_\Omega=1-\tau_w-\tau_l.$ Thus, \(\tau_w\) and \(\tau_l\) represent the probabilities that a randomly selected treatment subject wins or loses, respectively, against an independently selected control subject under the prespecified hierarchical comparison rule.

The corresponding unadjusted estimators are the two-sample \(U\)-statistics obtained by averaging over all \(n_1\times n_0\) treatment–control pairs,
\begin{equation}
\bar{\tau}_\nu = \frac{1}{n_1 n_0} \sum_{i:\, A_i = 1} \sum_{j:\, A_j = 0} \varphi_\nu(\mathcal{D}_i, \mathcal{D}_j),
\qquad \nu \in \{w, l\}.
\label{eq:u_statistics}
\end{equation} 

To characterize their first-order behavior, define the centered one-way conditional win--loss functions,
\begin{align}
    g_{\nu,T}(d)
    =
    \E[\varphi_\nu(d,\mathcal D_C)]-\tau_\nu, \quad \text{and} \quad
    g_{\nu,C}(d)
    =
    \E[\varphi_\nu(\mathcal D_T,d)]-\tau_\nu,
    \quad \nu\in\{w,l\}
    \label{eq:g_functions}
\end{align}
By construction, $\E[g_{\nu,T}(\mathcal{D}_T)]=\E[g_{\nu,C}(\mathcal{D}_C)]=0$. Here $g_{\nu,T}(d)$ is the centered expected contribution of a treated subject with HCE record $d$ compared against a random control, where $g_{\nu,C}(d)$ is the corresponding contribution of a control subject when compared with a random treated subject.  The first-order Hoeffding--H\'ajek projection of $\bar{\tau}_\nu-\tau_\nu$  
is \citep{vandervaart_asymptotic_1998}
$$
\Pi_\nu = \frac1{n_1}\sum_{i:\,A_i=1} g_{\nu,T}(\mathcal{D}_i) +\frac1{n_0}\sum_{j:\,A_j=0} g_{\nu,C}(\mathcal{D}_j).
$$ The projection $\Pi_\nu$ thus aggregates these one-way contributions across the two arms. The following result gives the corresponding two-sample Hoeffding decomposition.
\begin{lemma}[Hoeffding--H\'ajek projection]
\label{lem:hajek}
For $\nu \in \{w, l\}$, define the degenerate residual kernel $$r_\nu(d,d')=\varphi_\nu(d,d')-\tau_\nu-g_{\nu,T}(d)-g_{\nu,C}(d').$$ Then $\bar{\tau}_\nu-\tau_\nu= \Pi_\nu + R_{\nu,N}$ holds, where
$ R_{\nu,N}=\frac{1}{n_1 n_0}\sum_{i:\,A_i=1}\sum_{j:\,A_j=0}r_\nu(\mathcal{D}_i,\mathcal{D}_j).$ The residual kernel is degenerate in each argument. Since $\varphi_\nu$ is bounded and $n_1/N\to e\in(0,1)$, $R_{\nu,N}=\Op(N^{-1})$, and hence $\sqrt{N}\,R_{\nu,N}=\op(1)$.
\end{lemma}

Consequently, the first-order behavior of the pairwise U-statistic is completely characterized by the subject-level one-way projection terms.

\begin{corollary}[Influence function for unadjusted win and loss probabilities]\label{cor:if}
For \(\nu\in\{w,l\}\),
\[\sqrt{N} \left(\bar\tau_\nu-\tau_\nu\right)
=
\frac{1}{\sqrt{N}}\sum_{i=1}^N\psi_\nu(\mathcal O_i)+o_p(1),
\]
where the subject-level IF is 
\[
\psi_\nu(\mathcal O_i)=\frac{A_i}{e}\,g_{\nu,T}(\mathcal D_i) +\frac{1-A_i}{1-e}\,g_{\nu,C}(\mathcal D_i).
\] The IF satisfies $\E\{\psi_\nu(\mathcal O_i)\}=0$ and has finite variance.
\end{corollary} 
Proofs of Lemma \ref{lem:hajek} and Corollary \ref{cor:if} are given in the Supplementary Material. 

To extend the result from the elementary win and loss probabilities to commonly used win measures, stack the elementary win and loss probabilities and influence functions as
\[
\boldsymbol\tau=(\tau_w,\tau_l)^\top,
\qquad
\bar{\boldsymbol\tau}=(\bar\tau_w,\bar\tau_l)^\top,
\qquad
\boldsymbol\psi(O)=(\psi_w(O),\psi_l(O))^\top.
\]
The win measures considered here are smooth transformations of the common win-loss probability vector,
\[
\eta_m
=
h_m(\boldsymbol\tau),
\qquad
\bar\eta_m
=
h_m(\bar{\boldsymbol\tau}),
\qquad
m\in\mathcal M
=
\{\mathrm{NB},\mathrm{LWR},\mathrm{LWO}\},
\]
where NB, LWR, and LWO denote the net benefit, log win ratio, and log win odds, respectively. Letting $\Delta=\tau_w-\tau_l$ denotes the probabilities difference, then
\[
h_{\mathrm{NB}}(\boldsymbol\tau)=\Delta,\quad
h_{\mathrm{LWR}}(\boldsymbol\tau)=\log\tau_w-\log\tau_l,\quad
h_{\mathrm{LWO}}(\boldsymbol\tau)=\log(1+\Delta)-\log(1-\Delta).
\]
We assume $h_m$ is continuously differentiable at the true value of $\mathbf \tau$; in particular
$\tau_w,\tau_l>0$ for LWR and $|\Delta|<1$ for LWO. Stacking Corollary~\ref{cor:if} gives
$$
\sqrt N(\bar{\boldsymbol\tau}-\boldsymbol\tau)
=\frac1{\sqrt N}\sum_{i=1}^N\boldsymbol\psi(\mathcal O_i)+o_p(1),
$$
and the multivariate delta method yields
\begin{equation}
    \sqrt N(\bar\eta_m-\eta_m)
    =\frac1{\sqrt N}\sum_{i=1}^N\psi_{m}(\mathcal O_i)+o_p(1),
    \qquad \text{where} \, \,
    \psi_{m}(\mathcal O_i)=\nabla h_m(\boldsymbol\tau)^\top\boldsymbol\psi(\mathcal O_i).
\end{equation}
Thus, each win measure $m$ inherits its IF elementary vector $\boldsymbol\psi(\mathcal O_i)$. The corresponding estimands, gradients, and influence functions are summarized in Table ~\ref{tab:if_summary}.
\begin{table}[h]
\centering
\small
\caption{Unadjusted estimands and IF representations for all win measures.}
\label{tab:if_summary}
\renewcommand{\arraystretch}{2.2}
\begin{tabular}{lcccc}
\toprule
\textbf{Measure $m$}  & \textbf{Estimand $\eta_m$} & \textbf{Unadj. Estimator $\bar\eta_m$}
  & \textbf{Gradient $\nabla h_m(\boldsymbol\tau)^\top$} & \textbf{IF $\psi_{m}$}\\
\midrule
NB
  & $\Delta = \tau_w-\tau_l$
  & $\bar\Delta=\bar\tau_w-\bar\tau_l$
  & $(1,\,-1)$
  & $\psi_w-\psi_l$ \\
LWR
  & $\log\tau_w-\log\tau_l$
  & $\log\bar\tau_w-\log\bar\tau_l$
  & $\bigl(\frac{1}{\tau_w},\,-\frac{1}{\tau_l}\bigr)$
  & $\dfrac{\psi_w}{\tau_w}-\dfrac{\psi_l}{\tau_l}$ \\
LWO
  & $\log\dfrac{1+\Delta}{1-\Delta}$
  & $\log\dfrac{1+\bar\Delta}{1-\bar\Delta}$
  & $\bigl(\frac{2}{1-\Delta^2},\,-\frac{2}{1-\Delta^2}\bigr)$
  & $\dfrac{2}{1-\Delta^2}\bigl(\psi_w-\psi_l\bigr)$ \\
\bottomrule
\end{tabular}
\end{table}

\subsection{Variance Estimation and Unadjusted Test Statistics}
\label{subsec:unadj-test}

Because all win measures $m$ are smooth functions of the same win--loss probability vector $\boldsymbol\tau$, a single estimate of probability IF $\boldsymbol\psi(\mathcal O_i)$ can be used for variance estimation and inference across all measures. First, we estimate the one-way projection functions of \eqref{eq:g_functions} by their empirical two-sample analogues. For $\nu\in\{w,l\}$, we have
\begin{equation}
\label{eq:ghat-functions}
\widehat g_{\nu,T}(\mathcal D_i)
=
\frac1{n_0}\sum_{k:A_k=0}
\varphi_\nu(\mathcal D_i,\mathcal D_k)
-\bar\tau_\nu,
\quad
\widehat g_{\nu,C}(\mathcal D_i)
=
\frac1{n_1}\sum_{k:A_k=1}
\varphi_\nu(\mathcal D_k,\mathcal D_i)
-\bar\tau_\nu.
\end{equation}
The estimated IF scores for the win--loss probabilities $\tau_\nu$ are then,
\begin{equation}
\label{eq:estimated-elementary-if}
\widehat\psi_{\nu,i} = \frac{A_i}{e}\widehat g_{\nu,T}(\mathcal D_i) + \frac{1-A_i}{1-e}\widehat g_{\nu,C}(\mathcal D_i), \qquad \text{with} \,\,
\widehat{\boldsymbol\psi}_i = (\widehat\psi_{w,i},\widehat\psi_{l,i})^\top .
\end{equation}
By construction, \(N^{-1}\sum_{i=1}^N\widehat{\boldsymbol\psi}_i=\mathbf 0\), so their empirical covariance is 
$ \widehat{\boldsymbol\Sigma} = \frac1N\sum_{i=1}^N\widehat{\boldsymbol\psi}_i \widehat{\boldsymbol\psi}_i^\top.$
By the delta method, for each $m\in\mathcal M$,
\begin{equation}\label{eq:var-unadj}
\widehat{\mathrm{Var}}(\bar\eta_m) = \frac1N\,\nabla\widehat h_m^{\top}\,\widehat{\boldsymbol\Sigma}\,\nabla\widehat h_m, \qquad \text{with }\nabla\widehat h_m=\nabla h_m(\bar{\boldsymbol\tau}),
\end{equation}
which equals the empirical variance of the measure-specific scores $\widehat\psi_{m,i}=\nabla\widehat h_m^\top\widehat{\boldsymbol\psi}_i$.

To test the absence of a treatment effect, under $H_0:\eta_m=\eta_m^0$, we use the Wald statistic 
\begin{equation}\label{eq:Tunadj}
T_m=\frac{\bar\eta_m-\eta_m^0}{\sqrt{\widehat{\mathrm{Var}}(\bar\eta_m)}} \xrightarrow{d}\mathcal N(0,1)
\end{equation}
rejecting at two-sided level $\alpha$ when $|T_m|>z_{1-\alpha/2}$. 
All three measures share the no-effect null $\tau_w=\tau_l$, at which $\eta_m^0=0$ for every $m\in\mathcal M$.


\section{Covariate Adjusted Estimator and Theoretical Properties} \label{sec:Adjusted}

The unadjusted estimators of Section \ref{sec:unadjusted} ignore the baseline covariates $\mathbf{X}$. Under randomization where $A\perp \mathbf{X}$, covariate information can still be used to improve precision without altering the target marginal estimand. Let \(\mathbf X\) be a prespecified, fixed-dimensional vector of baseline covariates with finite and non-singular covariance matrix \(\boldsymbol\Sigma_X=\mathrm{Var}(\mathbf X)\). Throughout this section, the restriction horizon \(S\) is fixed as in Section \ref{sec:unadjusted}, and its dependence is suppressed for notational simplicity. 


\subsection{Model-assisted augmentation and assumption-lean validity}
Following \cite{zhang2025unified}, consider the augmented estimator of the elementary win--loss probability vector 
\begin{equation}\label{eq:adj_tau}
   \hat{\boldsymbol\tau}(\boldsymbol b)
   =\bar{\boldsymbol\tau} - \frac1N\sum_{i=1}^N (A_i-e)\,\boldsymbol b(\mathbf X_i),
\end{equation}
where $e=\Pr(A=1)$ is known by design and $\boldsymbol b(\mathbf X)=\left(b_w(\mathbf X), b_l(\mathbf X)\right)^\top$ is any square-integrable augmentation. The choice $\boldsymbol b\equiv\boldsymbol 0$ recovers the unadjusted estimator $\bar{\boldsymbol\tau}$. Randomization gives $\E[(A-e)\boldsymbol b(\mathbf X)] =\boldsymbol 0,$ any fixed augmentation $\boldsymbol b (\mathbf X)$ preserves consistency for the same marginal parameter \(\boldsymbol\tau\). 
Combining this mean-zero augmentation with Corollary~\ref{cor:if} gives 
\begin{equation}
    \sqrt N \left\{\hat{\boldsymbol\tau}(\boldsymbol b) - \boldsymbol\tau \right \} = \frac1{\sqrt N} \sum_{i=1}^N \left[ \boldsymbol\psi(\mathcal O_i) - (A_i-e)\boldsymbol b(\mathbf X_i) \right] +o_p(1). 
\end{equation} 
For a win measure \(m\), define augmented estimator of $\eta_m$ as $ \hat\eta_m(\boldsymbol b) = h_m\{\hat{\boldsymbol\tau}(\boldsymbol b)\}.$ 
The delta method gives 
\begin{equation}
    \sqrt N \left\{\hat\eta_m(\boldsymbol b)-\eta_m\right\} = \frac1{\sqrt N} \sum_{i=1}^N \left[\psi_m(\mathcal O_i) - (A_i-e)\nabla h_m(\boldsymbol\tau)^\top \boldsymbol b(\mathbf X_i)\right] + o_p(1).     
\end{equation} 
Thus, the augmentation is \emph{assumption-lean}: correctness of the working augmentation $\boldsymbol b(\mathbf X)$ is not required for consistency of the time-restricted marginal win estimand.

\subsection{Optimal augmentation as an efficiency benchmark}\label{subsec:optimal}
For a fixed vector augmentation \(\boldsymbol b\), the asymptotic variance of \(\hat\eta_m(\boldsymbol b)\) is
\begin{equation}\label{eq:var-adj-general}
    \sigma^2_m(\boldsymbol{b}) = \E \left[ \{\psi_m(\mathcal O) - (A-e)\nabla h_m(\boldsymbol\tau)^\top \boldsymbol b(\mathbf X) \}^2 \right].    
\end{equation}
A generic choice of \(\boldsymbol b\) need not improve efficiency compared to the unadjusted estimator; thus, \emph{assumption-lean} alone does not satisfy the \emph{guaranteed efficiency gain}. 
By the standard augmentation theory for randomized trials \citep{tsiatis2008covariate, zhang2008improving}, the variance-minimizing augmentation for the elementary IF vector is
\[
\boldsymbol b^{\mathrm{opt}}(\mathbf X)
=
E\{\boldsymbol\psi(\mathcal O)\mid A=1,\mathbf X\}
-
E\{\boldsymbol\psi(\mathcal O)\mid A=0,\mathbf X\}.
\]
Since \(\psi_m(\mathcal O)=\nabla h_m(\boldsymbol\tau)^\top\boldsymbol\psi(\mathcal O)\), the scalar augmentation induced by this vector, $\nabla h_m(\boldsymbol\tau)^\top\boldsymbol b^{\mathrm{opt}}(\mathbf X),$ is simultaneously the variance-minimizing augmentation for every \(m\). Therefore, the same elementary vector augmentation is optimal for all $\eta_m$. At the optimum, orthogonality of the residual IF and augmentation space gives 
$$\sigma_m^2(\boldsymbol b^{\mathrm{opt}}) = \Var\{\psi_m(\mathcal O)\} - \Var\left[ (A-e)\nabla h_m(\boldsymbol\tau)^\top \boldsymbol b^{\mathrm{opt}}(\mathbf X) \right] \leq \Var\{\psi_m(\mathcal O)\}.
$$ 
The oracle augmentation therefore provides the efficiency benchmark and cannot increase asymptotic variance. However, the form of $\boldsymbol b^{\mathrm{opt}}(\mathbf X)$ is generally unknown. Our second principle is stronger in practice: the working adjustment should retain this asymptotic \emph{no-loss efficiency} property even when the chosen working model does not contain the unrestricted optimum.

\subsection{Practical augmentation: linear projection and guaranteed efficiency gain} \label{subsec:practical}

We now restrict the augmentation to a common linear working space. Following the direct-risk minimization strategy of \citet{zhang2025unified}, for each \(\nu\in\{w,l\}\), define the population least-squares projection coefficient
\[
\beta_{\nu}^\star = \arg\min_{\beta\in\mathbb R^p} \E\left[ \left\{ \psi_\nu(O) - (A-e)\beta^\top\{\mathbf X - \E(\mathbf X)\} \right\}^2 \right], \qquad \nu\in\{w,l\}.
\]
For win measure \(m\), define $ \beta_{m}^\star = \partial_w h_m(\boldsymbol\tau)\beta_{w}^\star + \partial_l h_m(\boldsymbol\tau)\beta_{l}^\star.$ Because least-squares projection onto a common linear space is a linear operator, \(\beta_{m}^\star\) is exactly the coefficient obtained by directly projecting \(\psi_m\) onto \((A-e)\{\mathbf X-E(\mathbf X)\}\). Thus, a single pair of elementary projections, \((\beta_{w}^\star, \beta_{l}^\star)\), induces the linear projection for NB, LWR, and LWO simultaneously. The corresponding population residual variance is 
$ \sigma_{m,\mathrm{lin}}^2 = E[\{ \psi_m(O) - (A-e) \beta_{m}^{\star\top} \{\mathbf X-E(\mathbf X)\}\}^2].$ By orthogonality of the least-squares projection,  
$$
\Var\{\psi_m(O)\} - \sigma_{m,\mathrm{lin}}^2 = e(1-e) \beta_{m}^{\star\top} \boldsymbol\Sigma_X \beta_{m}^\star \ge 0. 
$$
Thus, the linear projection provides simultaneous asymptotic \emph{guaranteed efficiency gain} for NB, LWR, and LWO, without requiring the linear working model to be correctly specified.

In the trial sample, replace \(\E(\mathbf X)\) by \(\bar{\mathbf X}\), and replace \(\psi_\nu\) by \(\widehat\psi_{\nu,i}\) in \eqref{eq:estimated-elementary-if}. We estimate $\hat\beta_\nu = \arg\min_{\beta\in\mathbb R^p} \sum_{i=1}^N[\widehat\psi_{\nu,i} - (A_i-e)\beta^\top(\mathbf X_i-\bar{\mathbf X})]^2.$ Provided that the empirical Gram matrix is nonsingular, this yields the closed-form solution
\begin{equation}\label{eq:linear-beta}
\hat\beta_\nu = \Big\{ \sum_{i=1}^N (A_i-e)^2 (\mathbf X_i-\bar{\mathbf X}) (\mathbf X_i-\bar{\mathbf X})^\top \Big\}^{-1} \Big\{ \sum_{i=1}^N (A_i-e) (\mathbf X_i-\bar{\mathbf X}) \widehat\psi_{\nu,i} \Big\}.
\end{equation}
The fitted linear augmentation for component \(\nu\) is $ \hat b_\nu(\mathbf X_i) = \hat\beta_\nu^\top(\mathbf X_i-\bar{\mathbf X}).$  Substituting this fitted augmentation into \eqref{eq:adj_tau} gives the linearly adjusted win--loss probability estimator
\begin{equation}\label{eq:linear-adj-tau}
\hat\tau_\nu^{\mathrm{lin}} = \bar\tau_\nu - \frac1N \sum_{i=1}^N (A_i-e) \hat\beta_\nu^\top(\mathbf X_i-\bar{\mathbf X}) = \bar\tau_\nu - \frac{n_1n_0}{N^2} \hat\beta_\nu^\top (\bar{\mathbf X}_1-\bar{\mathbf X}_0), \qquad \nu\in\{w,l\},
\end{equation} 
where the second equality follows from sample centering: $ \frac1N \sum_{i=1}^N (A_i-e)(\mathbf X_i-\bar{\mathbf X}) = \frac{n_1n_0}{N^2}(\bar{\mathbf X}_1-\bar{\mathbf X}_0).$ Thus the linear augmentation removes the component of $\bar\tau_\nu$ that is explained by chance $\mathbf{X}$ imbalance between the randomized arms. Define linearly adjusted win--loss probability estimator vector as $\widehat{\boldsymbol\tau}^{\mathrm{lin}} = (\hat\tau_w^{\mathrm{lin}}, \hat\tau_l^{\mathrm{lin}})^\top$, and obtain each linearly adjusted win measure via 
\begin{equation}\label{eq:eta_adj}
    \hat\eta_m^{\mathrm{lin}} = h_m(\widehat{\boldsymbol\tau}^{\mathrm{lin}}).    
\end{equation}
Thus, $\bar{\boldsymbol\tau}$ is adjusted once,  and then obtains all win measures $\hat\eta_m^{\mathrm{lin}}$ from the same $\widehat{\boldsymbol\tau}^{\mathrm{lin}}$. 

\subsection{Linearly adjusted variance and test statistics}
\label{subsec:adj-test}
Denote the fitted vector augmentation as $ \widehat{\boldsymbol b}(\mathbf X_i) = \{ \hat\beta_w^\top(\mathbf X_i-\bar{\mathbf X}), \hat\beta_l^\top(\mathbf X_i-\bar{\mathbf X}) \}^\top.$ The estimated linearly adjusted elementary IF score vector is $$ 
\widehat{\boldsymbol\psi}^{\mathrm{lin}}_i = \widehat{\boldsymbol\psi}_i - (A_i-e)\widehat{\boldsymbol b}(\mathbf X_i).
$$ 
Let $ \bar{\widehat{\boldsymbol\psi}}^{\mathrm{lin}} = N^{-1}\sum_{i=1}^N \widehat{\boldsymbol\psi}^{\mathrm{lin}}_i$ and define $ \widehat{\boldsymbol\Sigma}^{\mathrm{lin}} = N^{-1} \sum_{i=1}^N ( \widehat{\boldsymbol\psi} ^{\mathrm{lin}}_i - \bar{\widehat{\boldsymbol\psi}} ^{\mathrm{lin}} ) ( \widehat{\boldsymbol\psi}^{\mathrm{lin}}_i - \bar{\widehat{\boldsymbol\psi}}^{\mathrm{lin}})^\top.$ 
For a win measure \(m\), the plug-in variance estimator is
\begin{equation} \label{eq:var-adj} 
\widehat{\Var}(\hat\eta_m^{\mathrm{lin}}) = \frac1N \widehat\nabla h_m^{\mathrm{lin}\top} \widehat{\boldsymbol\Sigma}^{\mathrm{lin}} \widehat\nabla h_m^{\mathrm{lin}}, \qquad m\in\mathcal M,
\end{equation}
where we denote $ \widehat\nabla h_m^{\mathrm{lin}} = \nabla h_m (\widehat{\boldsymbol\tau} ^{\mathrm{lin}}).$

For \(H_0:\eta_m=\eta_m^0\), the linearly adjusted Wald test statistic is 
\begin{equation}\label{eq:Tadj}
    T_m^{\mathrm{lin}} = \frac{ \hat\eta_m^{\mathrm{lin}}-\eta_m^0 }{ \sqrt{\widehat{\Var}(\hat\eta_m^{\mathrm{lin}})}} \xrightarrow{d} N(0,1).
\end{equation} 
A two-sided level-\(\alpha\) test rejects when $\left|T_m^{\mathrm{lin}}\right|>z_{1-\alpha/2}.$ Confidence intervals for LWR and LWO are formed on the log scale; if WR or WO is reported, the interval endpoints are exponentiated. 

\subsection{Practical implementation algorithm}\label{sec:algorithm}
\begin{figure}[h]
    \centering
    \includegraphics[width=0.95\linewidth]{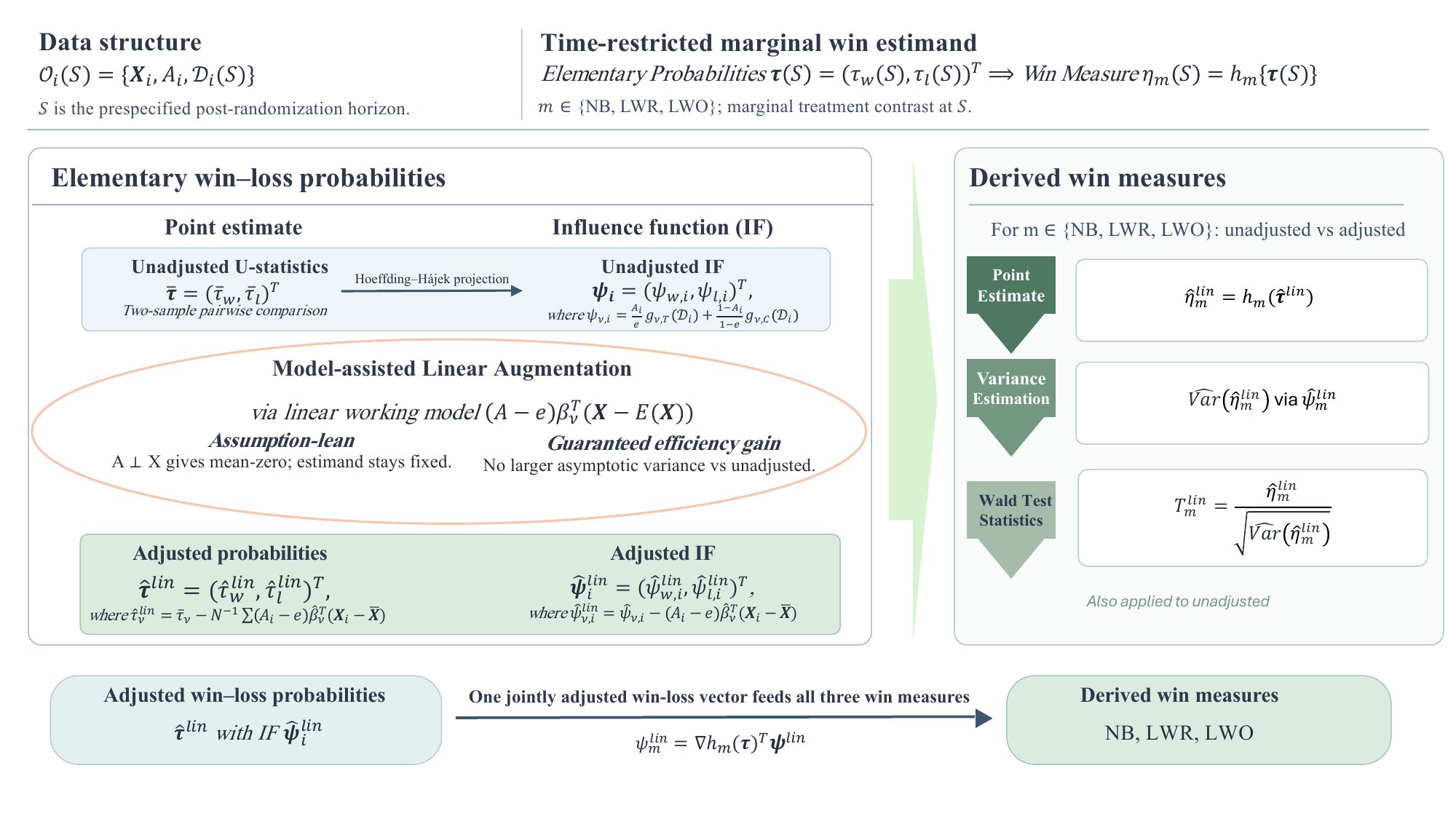}
    \caption{Schematic overview of the proposed covariate-adjustment algorithm for time-restricted marginal win statistics.}
    \label{fig:flow}
\end{figure}

For practical implementation, the proposed covariate-adjustment procedure is summarized in Figure~\ref{fig:flow} and the following six-step algorithm.

\begin{enumerate}[label=Step \arabic*., nosep]
\item \emph{Construct the fixed-horizon pairwise win--loss kernels.}
Fix the restriction horizon \(S\), construct the component-specific comparison rules (Table \ref{tab:comparison_rules}), and compute \(\varphi_w(\mathcal D_i,\mathcal D_j)\) and \(\varphi_l(\mathcal D_i,\mathcal D_j)\) according to \eqref{eq:kernel_functions} for every treatment-control pair.

\item \emph{Estimate the unadjusted win--loss probabilities and win measure.}
Compute unadjusted win--loss probability estimators $\bar{\boldsymbol\tau}=(\bar\tau_w,\bar\tau_l)^\top$ from \eqref{eq:u_statistics}, and then obtain the unadjusted win-measure estimators $\bar\eta_m=h_m(\bar{\boldsymbol\tau})$ for all $m\in\mathcal M$.

\item \emph{Estimate the elementary IF scores.}
Estimate the one-way placement functions $\widehat g_{\nu,T}$ and $\widehat g_{\nu,C}$, for $\nu\in\{w,l\}$, via \eqref{eq:ghat-functions}. Then construct the estimated elementary IF scores $\widehat\psi_{\nu,i} = \frac{A_i}{e}\widehat g_{\nu,T}(\mathcal D_i) + \frac{1-A_i}{1-e}\widehat g_{\nu,C}(\mathcal D_i),$ from \eqref{eq:estimated-elementary-if} and stack them as $\widehat{\boldsymbol\psi}_i = (\widehat\psi_{w,i},\widehat\psi_{l,i})^\top$.

\item \emph{Fit the elementary linear augmentations.}
Fit linear projections $ \hat\beta_\nu = \arg\min_{\beta\in\mathbb R^p} \sum_{i=1}^N [ \widehat\psi_{\nu,i} - (A_i-e)\beta^\top(\mathbf X_i-\bar{\mathbf X})]^2,\, $\(\nu\in\{w,l\}\), with closed-form solution given in \eqref{eq:linear-beta}. Set $ \widehat b_\nu(\mathbf X_i) = \hat\beta_\nu^\top(\mathbf X_i-\bar{\mathbf X}),$ and $ \widehat{\boldsymbol b}(\mathbf X_i) = \{\widehat b_w(\mathbf X_i), \widehat b_l(\mathbf X_i)\}^\top.$

\item \emph{Adjust the elementary probabilities and transform to win measures.} Compute linearly adjusted win--loss probability vector $ \widehat{\boldsymbol\tau}^{\mathrm{lin}} = \bar{\boldsymbol\tau} - \frac1N \sum_{i=1}^N (A_i-e)\widehat{\boldsymbol b}(\mathbf X_i)$ from \eqref{eq:linear-adj-tau}. 
Then, obtain adjusted win measure $ \hat\eta_m^{\mathrm{lin}} = h_m (\widehat{\boldsymbol\tau}^{\mathrm{lin}})$ for all $m\in\mathcal M$. 

\item \emph{Estimate adjusted variances and conduct inference.}
Construct the estimated linearly adjusted elementary IF score as $\widehat{\boldsymbol\psi}^{\mathrm{lin}}_i = \widehat{\boldsymbol\psi}_i - (A_i-e) \widehat{\boldsymbol b}(\mathbf X_i),$ estimate their covariance matrix \(\widehat{\boldsymbol\Sigma}^{\mathrm{lin}}\), and compute $ \widehat{\Var}(\hat\eta_m^{\mathrm{lin}})$ from \eqref{eq:var-adj}. Report the corresponding adjusted Wald statistic $ T_m^{\mathrm{lin}} = \frac{\hat\eta_m^{\mathrm{lin}}-\eta_m^0} {\sqrt{\widehat{\Var}(\hat\eta_m^{\mathrm{lin}})}}$ and confidence interval. For LWR and LWO, confidence intervals are formed on the log scale and exponentiated for reporting WR and WO.
\end{enumerate}

\section{Simulation Studies}\label{sec:simulation}

We evaluated the finite-sample performance of the proposed linearly adjusted estimators $\hat\eta_m^{\mathrm{lin}}$ and their unadjusted counterparts $\bar\eta_m$ for NB, LWR, and LWO in this section. The simulations were designed to examine three aspects in finite sample size: (i) estimation and inferential validity without requiring correctly specified working models; (ii) efficiency gains and improvements in statistical power without requiring correctly specified working models; and (iii) the impact of component-specific prognostic information, covariate selection, and adjustment complexity. 

\subsection{Data-generating mechanism}\label{sec:sim_dgm}
We first generated two-arm randomized trials with $N\in\{200,600,800, 1200\}$ independent subjects.  Treatment was assigned as $A_i\sim\mathrm{Bernoulli}(e)$ independently of baseline covariates, with known randomization probability $e=1/2$.

For each subject $i$, we generated a latent Gaussian vector $\mathbf V_i=(V_{i1},\ldots,V_{i5})^\top \sim\mathcal N_5(\mathbf 0,\mathbf R)$, where $R_{jj}=1$ and $R_{jk}=0.2$ for $j\ne k$. Continuous covariates were defined as $X_{ij}=V_{ij}$ for $j\in\{1,3,5\}$, whereas binary covariates were generated as $X_{ij}=I\{V_{ij}>\Phi^{-1}(1-\pi_j)\}$ for $j\in\{2,4\}$, with $(\pi_2,\pi_4)=(0.50,0.35)$. Thus, the continuous covariates were standard normal, and the binary covariates had
prevalences of $0.50$ and $0.35$, respectively. To construct expanded working adjustment sets, we additionally generated $X_{i6},\ldots,X_{i15}$ independently from standard normal distributions, independently of $\mathbf V_i$. These additional covariates had no direct effects on any outcome component.

We consider a three-component ($Q=3$) mixed-type HCE $\mathcal D_i(S)=\{D_{1i}(S),D_{2i}(S), D_{3i}(S)\}$, comprising a terminal TTE outcome, a bounded integer-valued duration (e.g., ICU days), and a positive continuous measurement, in descending clinical priority.  The common post-randomization restriction horizon was $S=28$ days, with no loss-to-follow-up before $S$.  Component-specific linear predictors were defined as
\begin{equation}\label{eq:sim_linear_predictor}
\zeta_{qi} =\theta_{q0}+\alpha_q A_i+ \boldsymbol\beta_q^\top\mathbf X_i, \qquad \mathbf X_i=(X_{i1},\ldots,X_{i5})^\top \qquad q=1,2,3,
\end{equation}
where $\alpha_q$ denotes the component-specific treatment effect, and $\boldsymbol\beta_q$ represents the prognostic effects. 

The terminal event time $T_{1i}$ followed a Weibull proportional hazards model with linear predictor $\zeta_{1i}$ and was recorded as $D_{1i}(S)=\{\widetilde T_{1i}(S),\delta_{1i}(S)\}$, where $\widetilde T_{1i}(S)=\min(T_{1i},S)$ and $\delta_{1i}(S)=I(T_{1i}\le S)$. The baseline hazard was calibrated to yield approximately $20\%$ marginal event incidence by $S$ in control arm. The second component was obtained by rounding a Gaussian latent duration with mean $\zeta_{2i}$ and bounding it to $[0,S]$. The third component followed a log-normal distribution with log-scale mean $\zeta_{3i}$ and was observed at $S$ only among subjects who survived through $S$, i.e. $\delta_{1i}(S)=0$. Longer event times and smaller values of $D_2$ and $D_3$ were favorable. 

We specified partially overlapping prognostic effects across the three HCE components. $X_3$ and $X_4$ were prognostic for all three components, $X_2$ for both $D_1$ and $D_2$, and $X_1$ and $X_5$ exclusively for $D_1$ and $D_3$, respectively. To assess the plausibility of the simulated prognostic structure, we summarized both the individual covariate effects and the overall predictive characteristics of each component using all five prognostic covariates (Table~\ref{tab:sim_prognostic_effects}).

\begin{table*}[t]
\centering
\small
\caption{Component-specific prognostic effects and overall model characteristics under the data generating mechanism.}
\label{tab:sim_prognostic_effects}

\begin{tabular}{@{}llcccccl@{}}
\toprule
Component & Effect scale
& $X_1$ & $X_2$ & $X_3$ & $X_4$ & $X_5$
& Overall prognostic strength \\
\midrule
$D_1$ & Hazard ratio
& 1.70 & 1.25 & 1.40 & 1.40 & --
& C-index $=0.700$ \\
$D_2$ & Mean shift (SD units)
& -- & 0.50 & 0.30 & 0.80 & --
& $R^2=0.260$ \\
$D_3$ & Log-mean shift (SD units)
& -- & -- & 0.40 & 0.30 & 0.60
& $R^2=0.392$ \\
\bottomrule
\end{tabular}

\par\smallskip
\begin{minipage}{0.98\textwidth}
\footnotesize
\textit{Notes:}
Overall predictive performance was evaluated numerically using all five prognostic covariates in independent control-arm.
Harrell's C-index was calculated for $D_1$ using
horizon-restricted survival records, whereas
out-of-sample predictive $R^2$ was calculated for
$D_2$ and $\log D_3$ among subjects surviving
through $S$.
\end{minipage}
\end{table*}

Under $H_0$, all treatment coefficients $\alpha_q$ were zero, yielding $\eta_m=0$ for all three win measures. Under the $H_A$, we set $(\alpha_1,\alpha_2,\alpha_3)\approx \{\log(0.808), -0.656, \log(0.935)\}$, corresponding to a hazard ratio of $0.808$ for $D_1$, a mean reduction of $0.656$ on the latent duration scale for $D_2$, and a geometric mean ratio of $0.935$ for $D_3$. These effects favored treatment across all three components and were calibrated to yield moderate power. The corresponding true values of $\eta_m$ were approximated using $10^7$ independently generated treatment--control pairs. Treatment coefficients were held fixed across sample sizes and working adjustment sets.  Additional sampling details and comparison rules are provided in the Supplementary Material.

\subsection{Simulation configurations and performance metrics}
\label{sec:sim_scenarios}

To reflect uncertainty about covariate selection in practice, we considered seven working adjustment sets (Table~\ref{tab:sim_prognostic_sets}). These included covariates prognostic of all three outcome components (Shared),  covariates prognostic for a given component ($D_q$-prog), and those prognostic of any component (Union). We further augmented the Union set with five (Union+5) or ten (Union+10) independent noise covariates to investigate the finite-sample impact of including non-prognostic variables. The unadjusted estimator $\bar{\eta}_m$ served as the reference (Unadj).

To characterize how much component-specific prognostic information captured by each working covariate set, we numerically evaluated its out-of-sample predictive performance for all three components(Table~\ref{tab:sim_prognostic_sets}). Harrell's C-index assessed discrimination for $D_1$, while $R^2$ quantified the explained variation in $D_2$ and $\log D_3$ among subjects event-free through $S$. For example, the $D_1$-complete set achieved higher discrimination for $D_1$ than the $D_3$-complete set (C-index: $0.700$ vs.\ $0.646$), whereas the latter explained more variation in $\log D_3$ ($R^2$: $0.392$ vs.\ $0.195$).  These measures characterize component-specific prognostic performance rather than efficiency for the overall marginal win estimand for each working set.

\begin{table*}[h]
\centering
\small
\caption{Working adjustment sets and component-specific predictive performance.}
\label{tab:sim_prognostic_sets}

\begin{tabular}{@{}llrccc@{}}
\toprule
Label & Working set & $p$
& $D_1$: C-index
& $D_2$: $R^2$
& $\log D_3$: $R^2$ \\
\midrule
Unadj, $\bar{\eta}_m$ & None & 0
& -- & -- & -- \\
Shared & $X_3,X_4$ & 2
& 0.644 & 0.215 & 0.188 \\
$D_1$-prog & $X_1,\ldots,X_4$ & 4
& 0.700 & 0.260 & 0.195 \\
$D_2$-prog & $X_2,X_3,X_4$ & 3
& 0.652 & 0.260 & 0.191 \\
$D_3$-prog & $X_3,X_4,X_5$ & 3
& 0.646 & 0.216 & 0.392 \\
Union & $X_1,\ldots,X_5$ & 5
& 0.700 & 0.260 & 0.392 \\
Union+5 & $X_1,\ldots,X_{10}$ & 10
& \multicolumn{3}{c}{Same as Union} \\
Union+10 & $X_1,\ldots,X_{15}$ & 15
& \multicolumn{3}{c}{Same as Union} \\
\bottomrule
\end{tabular}

\par\smallskip
\begin{minipage}{0.98\textwidth}
\footnotesize
\textit{Notes:}
$p$ denotes the number of adjustment covariates.
\end{minipage}
\end{table*}

For each $N$, we generated $R=10{,}000$ Monte Carlo (MC) trial replicates, and evaluated $\bar{\eta}_m$ and all seven adjusted estimators within each replicates. We assessed empirical bias, empirical standard deviation (ESD), average estimated standard error (ASE), and the ASE/ESD ratio. Inferential performance was evaluated using coverage of nominal $95\%$ Wald confidence intervals and Type I error or power of two-sided Wald tests at the nominal $5\%$ significance level. MC uncertainty in coverage and rejection probabilities was quantified using 95\% intervals either. Efficiency gain was summarized by the empirical percent variance reduction (PVR) relative to $\bar\eta_m$, 
$$
\widehat{\mathrm{PVR}}_m = 100\left\{ 1-\frac{\mathrm{ESD}^2(\hat\eta_{m}^{\mathrm{lin}})} {\mathrm{ESD}^2(\bar\eta_m)} \right\}.
$$
 
\subsection{Simulation Results}\label{sec:sim_results}
We focus on LWR in the main text; corresponding results for NB and LWO are provided in the Supplementary Material. Across all working sets, the proposed linearly adjusted estimators $\hat{\eta}^{\mathrm{lin}}_m$ exhibited small empirical biases relative to their sampling variability, with the absolute bias remaining below $2.6\%$ of the ESD for all three win measures. Complete operating characteristics, including bias, ESD, ASE, and ${\mathrm{ASE}}/{\mathrm{ESD}}$, are reported in the Supplementary Material.
    
\begin{figure*}[h]
\centering
\includegraphics[width=0.9\textwidth]{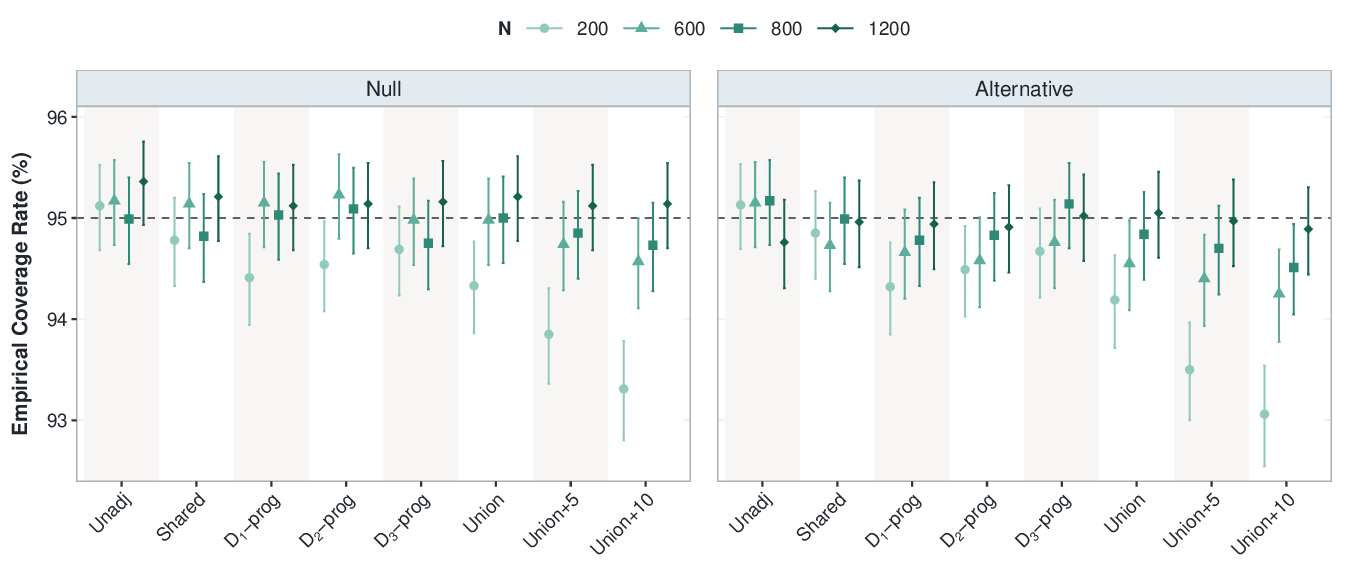}
\caption{Empirical coverage of nominal $95\%$ Wald confidence intervals for LWR under $H_0$ (left) and $H_A$ (right). Each point uses $R=10{,}000$ MC trials; error bars are pointwise $95\%$ Wilson MC confidence intervals.}
\label{fig:sim_coverage}
\end{figure*}

Empirical coverage rate of LWR in Figure \ref{fig:sim_coverage} exhibited two related patterns: it approached the nominal 95\% level as sample size $N$ increased, and it deteriorated as independent noise covariates were added to the Union adjustment set. The latter pattern was most pronounced at $N=200$. For example, under $H_A$, empirical coverage rate of LWR decreased from $94.19\%$ with Union to $93.50\%$ with Union+5 and $93.06\%$ with Union+10. Although undercoverage diminished with increasing sample size, it persisted for Union+10 at $N=600$ and $800$, with LWR coverage probabilities of $94.25\%$ and $94.51\%$, respectively. By $N=1200$, coverage approached the nominal level across all working adjustment sets. Similar patterns were observed under $H_0$; corresponding results for NB and LWO are presented in the Supplementary Material.

The coverage patterns in Figure \ref{fig:sim_coverage} were accompanied by ${\mathrm{ASE}} /{\mathrm{ESD}}$ below one, consistent with under-estimation of sampling variability contributing to the observed undercoverage of confidence interval (see Supplementary Figure and Tables). For LWR under $H_A$, ${\mathrm{ASE}}/ {\mathrm{ESD}}$ at $N=200$ decreased from $0.974$ with Union to $0.950$ with Union+5 and $0.928$ with Union+10. For Union+10, ${\mathrm{ASE}}/{\mathrm{ESD}}$ increased to $0.970$, $0.983$, and $0.991$ at $N=600$, $800$, and $1200$, respectively. Although the proposed approach is \emph{assumption-lean}, its IF-based analytical variance estimator and Wald confidence intervals are justified asymptotically, and their calibration may be imperfect in finite samples. This motivated a bootstrap variance estimation sensitivity analysis at $N=200$, presented in Section~\ref{sec:sim_bootstrap_variance}.

\begin{figure*}[h]
\centering
\includegraphics[width=\textwidth]{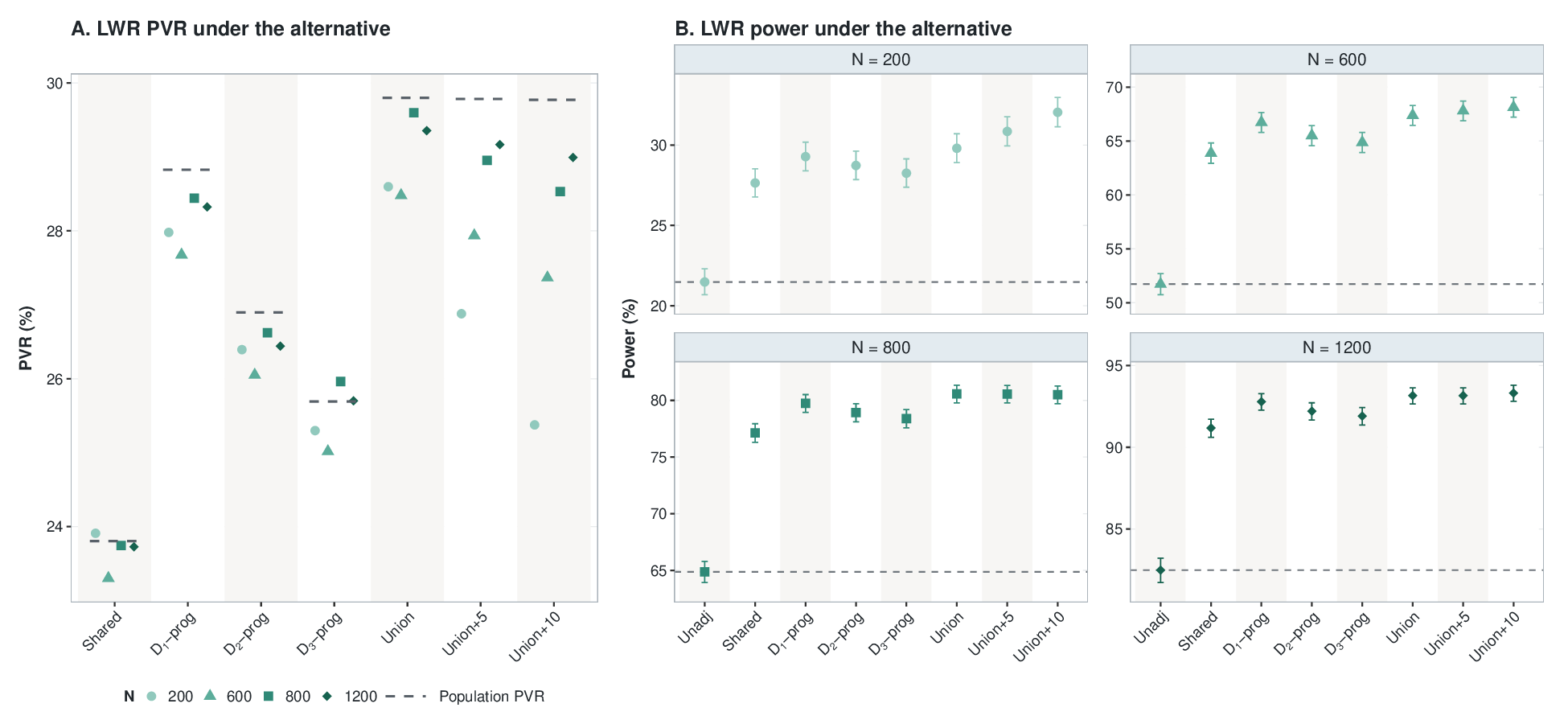}
\caption{Precision gain and power of LWR under $H_A$. Panel A compares empirical percent variance reduction (PVR) with the population benchmark (short dashed segments). Panel B shows power by sample size; horizontal dashed lines mark unadjusted power. Power error bars are pointwise $95\%$ Wilson Monte Carlo confidence intervals. }
\label{fig:lwr_pvr_power_by_working_set}
\end{figure*}

Figure~\ref{fig:lwr_pvr_power_by_working_set} presents precision gains and power for LWR under $H_A$. Panel~A provides empirical evidence consistent with the asymptotic \emph{guaranteed efficiency gain} relative to $\bar{\eta}_m$. Every working adjustment set yielded a positive empirical PVR at each $N$, with gains ranging from $23.30\%$ to $29.60\%$. Thus, adjustment remained beneficial even when the working set captured only part of the available prognostic information. Union and its noise-included versions had the largest population PVR benchmarks of approximately $29.8\%$, although including additional noise covariates incurred a finite-sample efficiency cost. At $N=200$, empirical PVR decreased from $28.60\%$ with Union to $25.38\%$ with Union+10. This difference narrowed at $N=1,200$, where the corresponding values were $29.35\%$ and $28.99\%$. These finite-sample differences are compatible with a guarantee that concerns asymptotic variance.

Panel~B shows that adjusted estimators $\hat{\eta}^{\mathrm{lin}}_m$ with all working covariate set had higher power than $\bar{\eta}_m$ within each $N$. For example, at $N=800$, power increased from $64.87\%$ without adjustment to $80.57\%$ with Union. The higher power of the noise-augmented sets at $N=200$, however, should be interpreted alongside their underestimated standard errors and under-coverage described above. Because these power estimates were not calibrated to a common empirical Type~I error rate, they do not establish an additional efficiency benefit from including noise covariates.

\subsection{Finite-Sample Variance Estimation: A Bootstrap Sensitivity Analysis}
\label{sec:sim_bootstrap_variance}

Although the proposed approach is \emph{assumption-lean}, its analytical variance estimator and Wald inference rely on asymptotic approximations. Related covariate-adjustment studies have documented finite-sample Type I error inflation and investigated small-sample corrections or bootstrap-based variance estimation \citep{van2025combining, zhang2026bridging}. Motivated by the undercoverage observed at $N=200$, we compared the analytical IF-based variance estimator in \eqref{eq:var-adj} with bootstrap variance estimation using the same $R=10{,}000$ simulated trials under $H_0$, with $2{,}000$ bootstrap resamples per trial. Both methods used identical point estimates, differing only in variance estimation. Implementation details are provided in the Supplementary Material.

Figure~\ref{fig:sim_bootstrap_variance} shows that the improvement in variance calibration was more pronounced at $N=200$. For LWR with Union+10, bootstrap increased ASE/ESD from $0.922$ to $1.007$ and reduced Type~I error from $6.69\%$ to $4.66\%$. At $N=800$, analytical inference was already closer to nominal: the corresponding ASE/ESD ratios were $0.982$ and $1.004$, and Type~I error rates were $5.27\%$ and $4.73\%$. These results support that bootstrap variance estimation can mitigate finite-sample anti-conservatism when the working adjustment set is relatively large compared with the sample size. Corresponding NB and LWO results are reported in the Supplementary Material.

\begin{figure*}[h]
\centering
\includegraphics[width=\textwidth]{
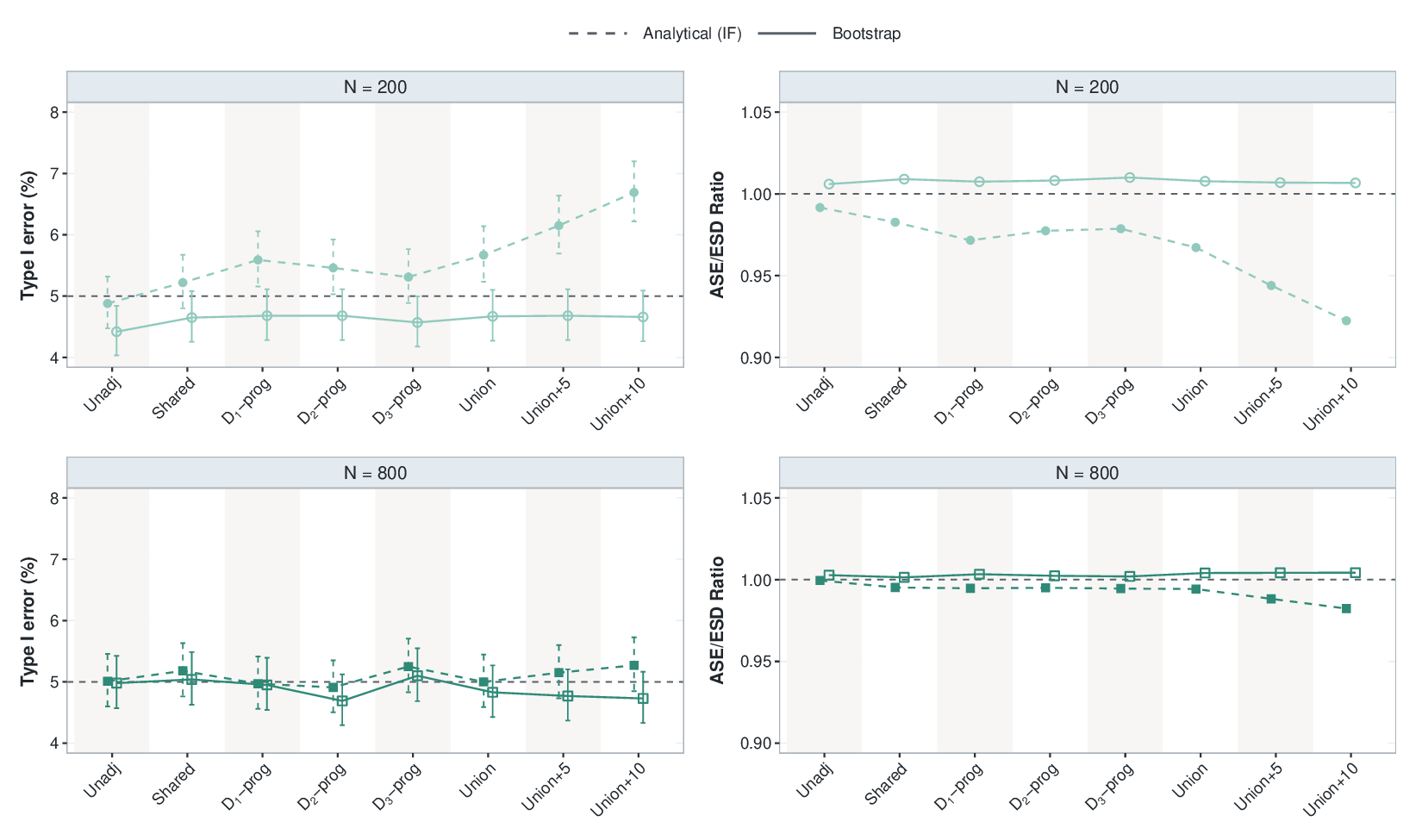}
\caption{Comparison of analytical IF-based and bootstrap variance estimation for LWR under $H_0$ at $N=200$ (top) and $N=800$ (bottom). The left column shows Type~I error; the right shows average estimated SE divided by empirical SD (ASE/ESD). Each sample size uses the same $R=10{,}000$ original trials, with $2{,}000$ bootstrap resamples per trial. Error bars are pointwise $95\%$ Wilson Monte Carlo confidence intervals.}
\label{fig:sim_bootstrap_variance}
\end{figure*}

\section{Real Data Analysis Example}\label{sec:realana}
We applied the proposed method to the double-blind, multicenter, parallel randomized clinical trial of electronic health record alerts for acute kidney injury (AKI) reported by \citet{wilson2021electronic}. The trial evaluated whether an automated electronic pop-up alert, accompanied by a link to an AKI order set, improved mortality, dialysis, and AKI progression within 14 days of randomization among adult inpatients with AKI. The analysis included all $N=6030$ randomized participants: $n_1=3059$ assigned to the alert ($A_i=1$) and $n_0=2971$ assigned to usual care ($A_i=0$).

The original primary outcome was the conventional composite that occurrence of AKI progression, dialysis, or death within 14 days of randomization. We used the same restriction horizon, $S=14$ days, but organized the three components hierarchically by clinical severity for illustration. Specifically, defined $D_{1i}(S)=I(\text{death by S})$, $D_{2i}(S)=I(\text{dialysis by S})$, and $D_{3i}(S)=I(\text{AKI progression by S})$, giving $\mathcal D_i(S)=\{D_{1i}(S),D_{2i}(S),D_{3i}(S)\}$. All components were adverse binary outcomes, so 0 was favorable. 

We compared the unadjusted estimator $\bar\eta_m$ with the proposed linear augmentation estimator $\widehat\eta_m^{\mathrm{lin}}$ described in Section~\ref{sec:algorithm}. The adjustment covariate set comprised the prespecified, fully observed baseline variables hospital, age over 90 years, sex, ICU status at randomization, Elixhauser comorbidity score, diabetes, and congestive heart failure. All three 14-day component outcomes and all baseline covariates
included in the adjustment set were fully observed in the analysis
dataset. All 6030 randomized participants were retained, without
imputation or complete-case deletion. 


\begin{table*}[h]
\centering
\footnotesize
\caption{Unadjusted and linearly adjusted estimates in the AKI alert trial. 
}
\label{tab:trial3_primary}
\setlength{\tabcolsep}{5pt}
\begin{tabular}{llrrrrrr}
\toprule
Measure & Method & Estimate & ASE & 95\% Wald CI   & $p$-value & PVR, \% \\
\midrule
\multirow{2}{*}{NB}
  & $\bar\eta_m$ & -0.0035 & 0.0106 & (-0.0243, 0.0173) & 0.742 & -- \\
  & $\widehat\eta_m^{\mathrm{lin}}$ & -0.0079 & 0.0101 & (-0.0277, 0.0119) & 0.434 & 9.2 \\
\addlinespace[2pt]
\multirow{2}{*}{LWR}
  & $\bar\eta_m$ & -0.0192 & 0.0583 & (-0.1335, 0.0950) & 0.742 & -- \\
  & $\widehat\eta_m^{\mathrm{lin}}$ & -0.0434 & 0.0555 & (-0.1522, 0.0653) & 0.434 & 9.4 \\
\addlinespace[2pt]
\multirow{2}{*}{LWO}
  & $\bar\eta_m$ & -0.0070 & 0.0212 & (-0.0486, 0.0346) & 0.742 & -- \\
  & $\widehat\eta_m^{\mathrm{lin}}$ & -0.0158 & 0.0202 & (-0.0554, 0.0238) & 0.434 & 9.2 \\
\bottomrule
\end{tabular}
\end{table*}

Table~\ref{tab:trial3_primary} shows little evidence of a treatment benefit on any win measure. Linear adjustment reduced the analytical standard error by around $5\%$
and the estimated variance by approximately $9\%$, but did not alter the substantive
conclusion. 

\section{Discussion}\label{sec:discussion}

We developed an \emph{assumption-lean} covariate adjustment framework for marginal win statistics over a prespecified follow-up horizon $S$, guided by two principles: \emph{assumption-lean} validity, so that consistency and asymptotically valid inference do not depend on correct specification of a working model, and \emph{guaranteed efficiency gain}, so that adjustment cannot increase asymptotic variance relative to the unadjusted estimator. By projecting subject-level influence functions onto baseline covariates, we establish these properties for marginal net benefit, log win ratio, and log win odds. Importantly, the prognostic value of covariates for individual components does not necessarily translate into equivalent efficiency gains for the overall marginal win estimand.

Our development assumes complete ascertainment of the outcomes required for hierarchical comparisons through $S$. This assumption is plausible to acute-care trials with relatively short, protocol-specified follow-up periods \citep{tavares2024dapagliflozin, andromeda2025personalized}, but may be restrictive with longer follow-up. Although the fixed-horizon estimand remains well defined, censoring and missing outcomes require additional assumptions and estimation strategies, such as inverse probability of censoring weighting or outcome modeling \citep{dong2020inverse,wang2023missing, mao2024defining}. Missing baseline covariates pose a distinct challenge: unlike missing outcomes, they may limit the prognostic information available for adjustment. Approaches such as baseline imputation may accommodate incomplete covariate information, although their implications for adjustment efficiency warrant further investigation \citep{benkeser2021improving}.  Importantly, fixed-horizon HCEs are not confined to short-term trials: EMPULSE and TRILUMINATE Pivotal evaluated hierarchical outcomes via win ratio over 90 days and one year, respectively \citep{voors2022sglt2,sorajja2023transcatheter}. Extending the proposed framework to accommodate incomplete data while preserving assumption-lean adjustment and asymptotic no-loss efficiency could therefore broaden its applicability to longer-term trials.

The efficiency guarantee is asymptotic and does not necessarily translate into improved finite-sample performance. In our simulations, adjustment for a small set of moderately to strongly prognostic covariates yielded substantial variance reduction. Adding up to ten independent non-prognostic covariates only modestly reduced empirical efficiency and did not eliminate the gains relative to unadjusted estimation, even at $N=200$. This limited loss may partly reflect the substantial prognostic information captured by the original adjustment set and the relatively low adjustment dimension. Whether similar benefits would persist with weaker prognostic relationships or larger adjustment sets remains uncertain. Nevertheless, the additional noise covariates led to greater underestimation of analytical standard errors and Type I error inflation in smaller samples, consistent with finite-sample concerns documented in previous studies previous studies \citep{tackney2023comparison, bannick2026covariate,
scheidegger2026covariate, zhang2026bridging}. Bootstrap variance estimation improved inferential calibration in the settings examined by providing a more conservative variance estimation \citep{tsiatis2008covariate, ZhangVarEst2026}. These findings highlight the importance of balancing prognostic information against adjustment complexity and motivate further investigation of small-sample inference.

Finally, incorporating prospective efficiency gains into sample size and power calculations for win statistics, including extensions of the FORSS framework \citep{zhang2026forss}, is a natural next step. Because the magnitude of these gains may be uncertain at the planning stage, group sequential and information-adaptive designs offer complementary opportunities to improve trial efficiency \citep{van2025combining}. 

\section*{Acknowledgment}
\section*{Data Availability}

The de-identified subject level data used in the real-data application are publicly available through the RCT Benchmark resource \citep{shao2026should, wilson2021electronic}. The R code implementing the proposed methods and reproducing the simulation study and real-data analysis is available at \url{https://github.com/BaoshanZZ/CovWin}.

\bibliographystyle{plainnat}
\bibliography{ref}

\newpage

\section{Supplementary Materiel}

\subsection{Proof of Lemma 1: Hoeffding--H\'ajek Projection} \label{App:Lemma1}
\begin{proof}
We prove the result for $\nu\in\{w,l\}$; the two cases are identical, and we suppress $\nu$ where no confusion arises. By the definition of the degenerate residual kernel
$r_\nu$,
\[
\varphi_\nu(d_i,d_j)-\tau_\nu =
g_{\nu,T}(d_i)+g_{\nu,C}(d_j)+r_\nu(d_i,d_j).
\]
Substituting this identity into the double average for $\bar\tau_\nu-\tau_\nu$ in gives
\begin{align*}
\bar\tau_\nu-\tau_\nu
=&\frac{1}{n_1 n_0}\sum_{i:\,A_i=1}\sum_{j:\,A_j=0}g_{\nu,T}(\mathcal{D}_i) +\frac{1}{n_1 n_0}\sum_{i:\,A_i=1}\sum_{j:\,A_j=0}g_{\nu,C}(\mathcal{D}_j)\\
&+\frac{1}{n_1 n_0}\sum_{i:\,A_i=1}\sum_{j:\,A_j=0}r_\nu(\mathcal{D}_i,\mathcal{D}_j).
\end{align*}
In the first term the inner sum over the $n_0$ control subjects contributes a factor $n_0$, reducing it to $n_1^{-1}\sum_{i:A_i=1}g_{\nu,T}(\mathcal{D}_i)$; symmetrically, the second term reduces to $n_0^{-1}\sum_{j:A_j=0}g_{\nu,C}(\mathcal{D}_j)$. This proves the exact
decomposition 
\begin{equation*}
   \bar{\tau}_\nu-\tau_\nu= \Pi_\nu + R_{\nu,N},
\label{eq:hajek}
\end{equation*} which is the two-sample Hoeffding decomposition
\citep{hoeffding_class_1948, lehmann_elements_1951}, with leading terms equal to the H\'ajek projection of $\bar\tau_\nu$ \citep[][Thm.~12.6]{vandervaart_asymptotic_1998}. 

The residual kernel $r_\nu$ is degenerate, where it has zero conditional mean in each argument. Using
$\E[\varphi_\nu(d,\mathcal{D}_j)\mid A_j=0]=g_{\nu,T}(d)+\tau_\nu$ (from the definition of
$g_{\nu,T}$) and the centering $\E[g_{\nu,C}(\mathcal{D}_j)\mid A_j=0]=0$, for fixed $d$,
\[
\E[r_\nu(d,\mathcal{D}_j)\mid A_j=0]
=\big(g_{\nu,T}(d)+\tau_\nu\big)-\tau_\nu-g_{\nu,T}(d)
 -\E[g_{\nu,C}(\mathcal{D}_j)\mid A_j=0]=0,
\]
and symmetrically $\E[r_\nu(\mathcal{D}_i,d)\mid A_i=1]=0$.

Let $\sigma_{r,\nu}^2=\E\{r_\nu(\mathcal{D}_i,\mathcal{D}_j)^2\mid A_i=1,A_j=0\}<\infty$,
which holds because the kernel is bounded, $0\le\varphi_\nu\le 1$, so that $r_\nu$ and the
projections in $g_{\nu, T}(d)$ and $g_{\nu, C}(d)$ are bounded. Expanding the variance,
\[
\Var(R_{\nu,N})=\frac{1}{n_1^2 n_0^2}
\sum_{i,k:\,A_i=A_k=1}\ \sum_{j,l:\,A_j=A_l=0}
\E\{r_\nu(\mathcal{D}_i,\mathcal{D}_j)\,r_\nu(\mathcal{D}_k,\mathcal{D}_l)\}.
\]
\begin{itemize}
    \item If $i\ne k$ and $j\ne l$, the four endpoints are independent and the two factors $\E\{r_\nu(\mathcal{D}_i,\mathcal{D}_j)\}$ and $\E\{r_\nu(\mathcal{D}_k,\mathcal{D}_l)\}$ have mean zero, so the expectation vanishes.
    \item If $i=k$ but $j\ne l$, condition on the shared $\mathcal{D}_i$ and apply degeneracy in the second argument; the cross expectation vanishes.
    \item  If $i\ne k$ but $j=l$, condition on the shared $\mathcal{D}_j$ and apply degeneracy in the first argument; it again vanishes.
\end{itemize}
  Hence only the $n_1 n_0$ terms with
$i=k$ and $j=l$ survive, giving
\[
\Var(R_{\nu,N})
=\frac{n_1 n_0}{n_1^2 n_0^2}\,\sigma_{r,\nu}^2
=\frac{\sigma_{r,\nu}^2}{n_1 n_0}.
\]
Since $n_1 n_0\asymp N^2$ under the stated rates, $R_{\nu,N}=\Op(N^{-1})$, and therefore
$\sqrt{N}\,R_{\nu,N}=\op(1)$.
\end{proof}

\subsection{Proof of Corollary \ref{cor:if}: Influence function for unadjusted win--loss probability $\bar{\tau}_\nu$}\label{App:Cor1}

\begin{proof}
Let $e_N=n_1/N$. The H\'ajek projection of $\bar{\tau}_\nu-\tau_\nu$ can be written as a single average over
all subjects,
\[
\Pi_\nu
=\frac1N\sum_{i=1}^N
\left\{
\frac{A_i}{e_N}\,g_{\nu,T}(\mathcal{D}_i)
+\frac{1-A_i}{1-e_N}\,g_{\nu,C}(\mathcal{D}_i)
\right\}.
\]
Subtracting the target i.i.d.\ sum, in which $e_N$ is replaced by its limit $e$, gives
\[
\Pi_\nu-\frac1N\sum_{i=1}^N\psi_\nu(O_i)
=\Big(\frac1{e_N}-\frac1e\Big)\frac1N\sum_{i=1}^N A_i\,g_{\nu,T}(\mathcal{D}_i)
+\Big(\frac1{1-e_N}-\frac1{1-e}\Big)\frac1N\sum_{i=1}^N (1-A_i)\,g_{\nu,C}(\mathcal{D}_i).
\]
Since $e_N\xrightarrow{p}e\in(0,1)$, both prefactors $\Big(\frac1{e_N}-\frac1e\Big)$ and $\Big(\frac1{1-e_N}-\frac1{1-e}\Big)$ are $o_p(1)$ by continuous mapping theorem. 

Moreover, \(g_{\nu,T}\) is bounded because \(0\le \varphi_\nu\le 1\), and hence
\(A g_{\nu,T}(\mathcal D)\) has finite variance. Therefore,
\[
\Var\Big\{
\frac1N\sum_{i=1}^N A_i g_{\nu,T}(\mathcal D_i)
\Big\}
=
\frac{1}{N}\Var\{A g_{\nu,T}(\mathcal D)\}
=O(N^{-1}).
\]
By Chebyshev inequality, 
\[
\frac1N\sum_{i=1}^N A_i g_{\nu,T}(\mathcal D_i)
=O_p(N^{-1/2}).
\]
The same argument gives for the control arm 
\[
\frac1N\sum_{i=1}^N (1-A_i)g_{\nu,C}(\mathcal D_i)
=O_p(N^{-1/2}).
\]
Consequently,
\[
\Pi_\nu-\frac1N\sum_{i=1}^N\psi_\nu(O_i)
=
o_p(1)O_p(N^{-1/2})+o_p(1)O_p(N^{-1/2})
=
o_p(N^{-1/2}).
\]

Combining this with the exact decomposition $\bar\tau_\nu-\tau_\nu=\Pi_\nu+R_{\nu,N}$ from
Lemma \ref{lem:hajek} and $R_{\nu,N}=\Op(N^{-1})=o_p(N^{-1/2})$ yields
$\bar\tau_\nu-\tau_\nu=N^{-1}\sum_{i=1}^N\psi_\nu(O_i)+o_p(N^{-1/2})$. Multiplying by
$\sqrt N$ gives the stated asymptotically linear representation.
\end{proof}

\subsection{Table 2: Influence function for all win measures}
\label{app:if-winmeasure}
By Corollary 2,
\[
\bar{\boldsymbol\tau}-\boldsymbol\tau
=
\frac1N
\sum_{i=1}^N
\boldsymbol\psi(O_i)
+
o_p(N^{-1/2}).
\]

Let \(\eta=h(\boldsymbol\tau)\) be any scalar win measure for which
\(h:\mathbb R^2\to\mathbb R\) is continuously differentiable in a neighborhood of
\(\boldsymbol\tau\), and let $\bar\eta=h(\bar{\boldsymbol\tau})$
denote its unadjusted plug-in estimator. A first-order Taylor expansion gives
\[
\bar\eta-\eta
=
\nabla h(\boldsymbol\tau)^\top
(\bar{\boldsymbol\tau}-\boldsymbol\tau)
+
o_p\left(\|\bar{\boldsymbol\tau}-\boldsymbol\tau\|\right).
\]
Since
\[
\bar{\boldsymbol\tau}-\boldsymbol\tau=O_p(N^{-1/2}),
\]
the Taylor remainder is \(o_p(N^{-1/2})\). Therefore,
\[
\bar\eta-\eta
=
\nabla h(\boldsymbol\tau)^\top
\left\{
\frac1N
\sum_{i=1}^N
\boldsymbol\psi(O_i)
\right\}
+
o_p(N^{-1/2}),
\]
or equivalently,
\[
\bar\eta-\eta
=
\frac1N
\sum_{i=1}^N
\psi_\eta(O_i)
+
o_p(N^{-1/2}),
\]
where
\[
\psi_\eta(O)
=
\nabla h(\boldsymbol\tau)^\top\boldsymbol\psi(O).
\]

We now record the resulting influence functions for the three win measures considered in
the main text.

First, for the net benefit (NB), we have
\[
h_{\mathrm{NB}}(\tau_w,\tau_l)=\tau_w-\tau_l,
\qquad
\nabla h_{\mathrm{NB}}(\boldsymbol\tau)=
\begin{pmatrix}
1\\
-1
\end{pmatrix}.
\]
Thus
\[
\psi_\mathrm{NB}(O)=\psi_w(O)-\psi_l(O).
\]

Second, for the log win ratio (LWR)
\[
\eta_{\mathrm{LWR}}=\log\tau_w-\log\tau_l,
\qquad
\bar\eta_{\mathrm{LWR}}=\log\bar\tau_w-\log\bar\tau_l,
\]
assume \(\tau_w>0\) and \(\tau_l>0\). Here
\[
h_{\mathrm{LWR}}(\tau_w,\tau_l)=\log\tau_w-\log\tau_l,
\qquad
\nabla h_{\mathrm{LWR}}(\boldsymbol\tau)=
\begin{pmatrix}
1/\tau_w\\
-1/\tau_l
\end{pmatrix}.
\]
Therefore,
\[
\psi_\mathrm{LWR}(O)
=
\frac{\psi_w(O)}{\tau_w}
-
\frac{\psi_l(O)}{\tau_l}.
\]

Third, for the log win odds (LWO), recall that
$
\tau_\Omega=1-\tau_w-\tau_l
$
and
$
WO
=
\frac{\tau_w+\frac12\tau_\Omega}
{\tau_l+\frac12\tau_\Omega}.
$
Let
$\Delta=\tau_w-\tau_l.
$
Thus the log win odds is
\[
\eta_{\mathrm{LWO}}
=
\log(1+\Delta)-\log(1-\Delta),
\]
with unadjusted estimator
\[
\bar\eta_{\mathrm{LWO}}
=
\log(1+\bar\Delta)-\log(1-\bar\Delta).
\]
Assume \(|\Delta|<1\). Define
\[
h_{\mathrm{LWO}}(\tau_w,\tau_l)
=
\log(1+\tau_w-\tau_l)
-
\log(1-\tau_w+\tau_l).
\]
Then
\[
\nabla h_{\mathrm{LWO}}(\boldsymbol\tau)
=
\begin{pmatrix}
\displaystyle \frac{2}{1-\Delta^2}\\[6pt]
\displaystyle -\frac{2}{1-\Delta^2}
\end{pmatrix}.
\]
Therefore,
\[
\psi_\mathrm{LWO}(O)
=
\frac{2}{1-\Delta^2}
\{\psi_w(O)-\psi_l(O)\}
=
\frac{2}{1-(\tau_w-\tau_l)^2}
\{\psi_w(O)-\psi_l(O)\}.
\]

Consequently, for each measure \(m\in\{\mathrm{NB},\mathrm{LWR},\mathrm{LWO}\}\),
\[
\sqrt N(\bar\eta_m-\eta_m)
=
\frac1{\sqrt N}
\sum_{i=1}^N
\psi_{\eta_m}(O_i)
+
o_p(1),
\]
with asymptotic variance
\[
\sigma_m^2
=
E\{\psi_{\eta_m}(O)^2\}.
\]

\subsection{Additional simulation details}\label{supp:sim_details}

This section supplies the sampling and computational details
not given in the main text. It complements the primary simulation results
already included in this Supplementary Material.

\paragraph{Outcome sampling and numerical specifications}
\label{supp:sim_sampling}
Let $\zeta_{qi}$ denote the raw-covariate predictors specified in the main
text. Independently of the covariates and treatment assignments, we drew
$\varepsilon_{2i},\varepsilon_{3i}\overset{\mathrm{iid}}{\sim}N(0,1)$,
independently of the random draws used to generate terminal event times.
All draws were independent across subjects. The component models were
\begin{align}
  T_{1i}\mid A_i,\mathbf X_i
  &\sim\mathrm{Weibull}(\text{shape}=k,\text{scale}=s_{1i}),
  \qquad k=1.2,\label{eq:supp_primary_tte}\\
  D_{2i}^{*}
  &=\zeta_{2i}+\sigma_2\varepsilon_{2i},
  \qquad \sigma_2=3.5,\label{eq:supp_primary_latent2}\\
  \log D_{3i}(S)
  &=\zeta_{3i}+\sigma_3\varepsilon_{3i},
  \qquad \sigma_3=0.45,\quad T_{1i}>S.
  \label{eq:supp_primary_log3}
\end{align}
The observed second component was obtained by rounding and bounding:
\[
  D_{2i}(S)
  =\min\{S,\max(0,\operatorname{round}(D_{2i}^{*}))\}.
\]
The quantities $\sigma_2$ and $\sigma_3$ are residual SDs on the latent second-component and log third-component scales, not marginal outcome SDs. For a subject with a terminal event by $S$, the third component was recorded as structurally unavailable, not imputed. 

The additional adjustment variables had exactly zero coefficients in all
three component models. Specifically,
$X_7\sim\mathrm{Bernoulli}(0.5)$ and
$X_9\sim\mathrm{Bernoulli}(0.3)$, while
$X_6,X_8,X_{10},\ldots,X_{15}$ were standard normal.
All ten noise variables were mutually independent and independent of the
five prognostic covariates, treatment assignments, and generation errors.

\paragraph{Hierarchical comparison rules}\label{supp:sim_comparison}
If exactly one subject experienced the terminal event by $S$, the survivor won. If both experienced the event, the subject with the later event time won. Equal terminal event times produced an overall tie, an event of probability zero under the continuous sampling model. Only pairs in which both subjects survived through $S$ proceeded to the lower-priority components. At the second component, we assume the subject with the smaller value won if the absolute difference exceeded two ($\epsilon_2=2$); otherwise comparison proceeded to the third component. At the third component, the smaller value won unless the larger-to-smaller ratio was at most $1.25$, in which case the pair was an overall tie. The latter criterion is equivalent to an absolute log-scale difference at most $\log(1.25)$. No structurally unavailable third-component measurement was required for any comparison. Each pair contributed exactly one win, loss, or tie.

\paragraph{Independent population diagnostics and benchmarks}
\label{supp:sim_population}

The prediction models underlying the working-set diagnostics were fitted
in $100{,}000$ independent control subjects and evaluated in a separate
control validation population of the same size. For each working set,
a Cox model predicted the first component, and linear models predicted
the observed second component and log third component among survivors.
Only the covariates in that working set were included.

Harrell's C-index used the observed horizon-restricted record
$\{\widetilde T_1(S),\delta_1(S)\}$, with larger risk scores indicating
greater hazard. For $D_2(S)$ and $\log D_3(S)$,
out-of-sample predictive $R^2$ was calculated as one minus
the mean squared prediction error divided by the
observed outcome variance, both evaluated among validation
controls surviving through $S$. 

Correlated covariates could provide
proxy information about omitted direct predictors.
Adding independent noise did not add population-level prognostic information,
although finite training and validation samples could yield small differences.
These component prediction models were not used in the win-statistic adjustment.

\paragraph{Bootstrap Variance Estimation}

We evaluated bootstrap variance estimation using the same $R=10{,}000$ simulated trials under $H_0$ at $N=200$. For each trial, we generated $B=2{,}000$ bootstrap samples by drawing $N$ subjects with replacement from the pooled trial sample. Each subject's treatment assignment, baseline covariates, and outcome data were resampled jointly. Treatment-group sizes could therefore vary across resamples, while the known allocation probability remained fixed at $e=0.5$.

For each working adjustment set, the complete estimation procedure was repeated within every bootstrap sample, including estimation of win and loss probabilities, placements, influence-function contributions, covariate centering, and linear adjustment coefficients. The resulting estimates were then transformed to NB, LWR, and LWO. For estimand $m$, the bootstrap standard error was
\[
\widehat{\mathrm{SE}}_{\mathrm{boot},m}
=
\left[
\frac{1}{B-1}
\sum_{b=1}^{B}
\left\{
\hat{\eta}^{*(b)}_m-\bar{\eta}^{*}_m
\right\}^{2}
\right]^{1/2},
\qquad
\bar{\eta}^{*}_m
=
\frac{1}{B}\sum_{b=1}^{B}\hat{\eta}^{*(b)}_m.
\]

\subsubsection*{Supplementary Simulation Results}

\begin{figure*}[h]
    \centering
    \includegraphics[width=0.8\linewidth]{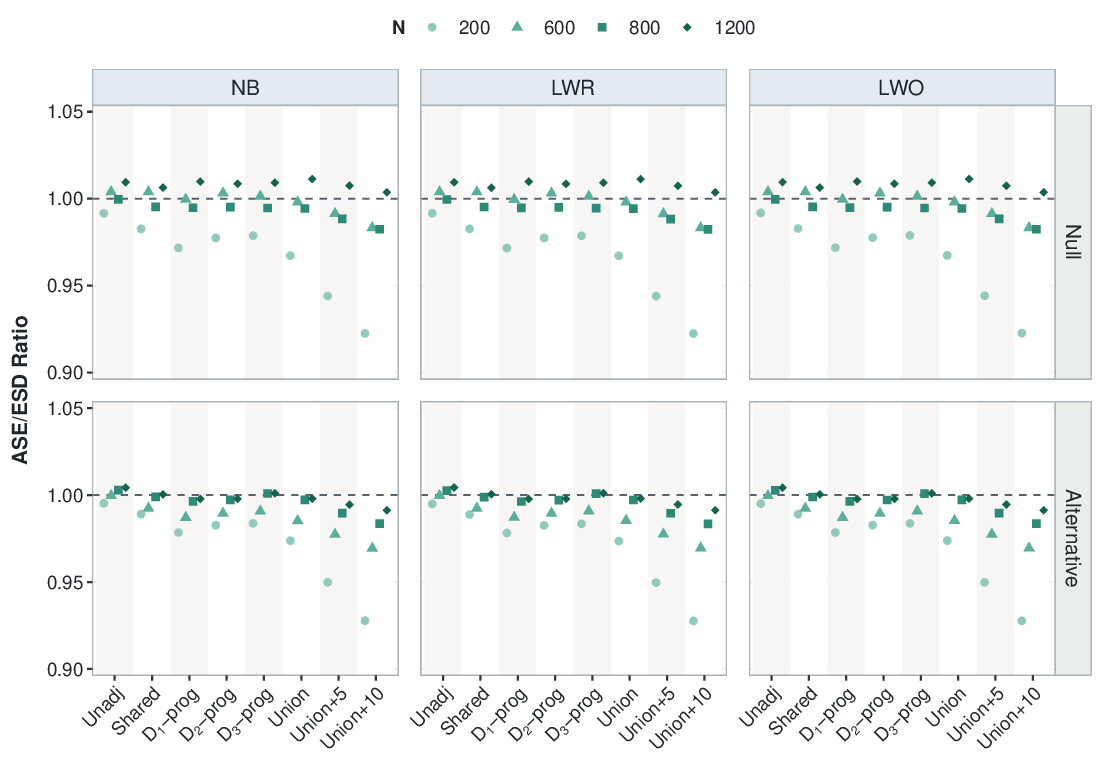}
    \caption{Standard-error calibration across working adjustment sets and sample sizes.}
    \label{fig:placeholder}
\end{figure*}

\begin{figure*}[h]
    \centering
    \includegraphics[width=0.9\linewidth]{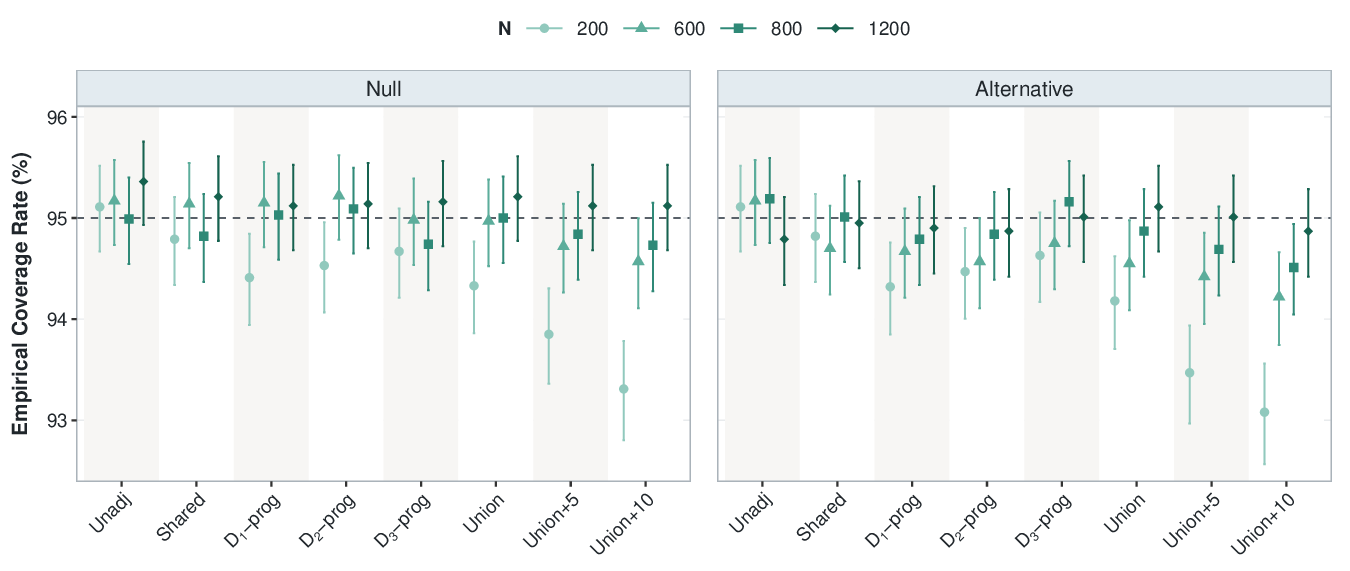}
    \caption{Empirical coverage of nominal $95\%$ Wald confidence intervals for LWO under $H_0$ (left) and $H_A$ (right). Each point uses $R=10{,}000$ MC trials; error bars are pointwise $95\%$ Wilson MC confidence intervals.}
    \label{fig:placeholder}
\end{figure*}

\begin{figure*}[h]
    \centering
    \includegraphics[width=0.9\linewidth]{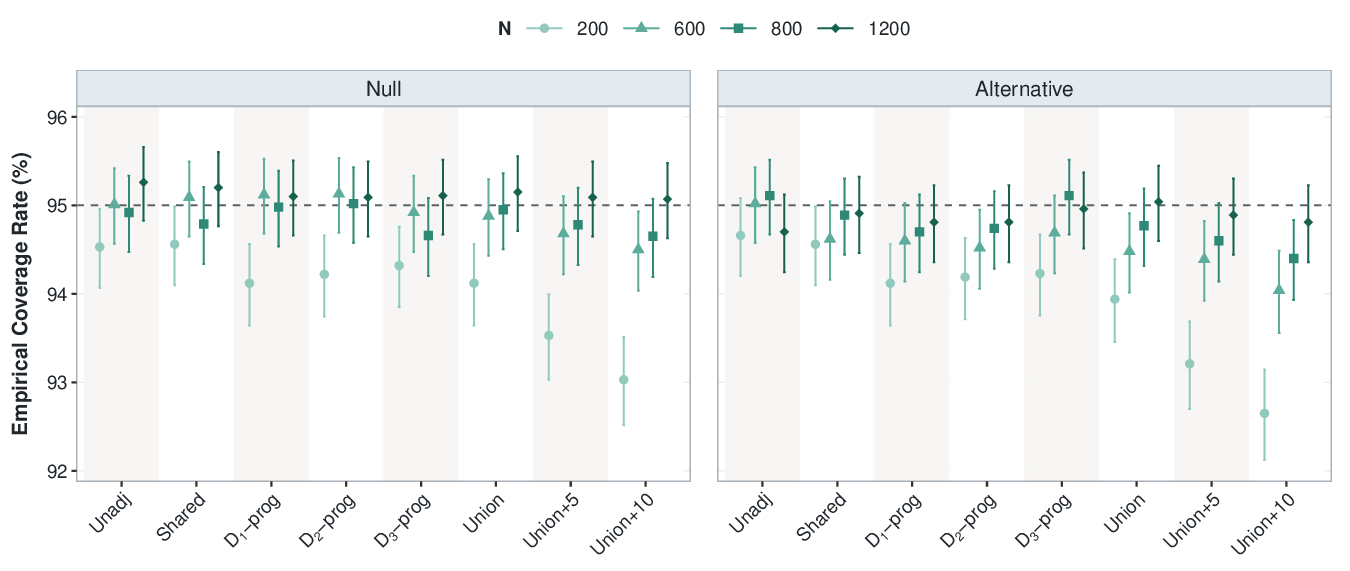}
    \caption{Empirical coverage of nominal $95\%$ Wald confidence intervals for NB under $H_0$ (left) and $H_A$ (right). Each point uses $R=10{,}000$ MC trials; error bars are pointwise $95\%$ Wilson MC confidence intervals.}
    \label{fig:placeholder}
\end{figure*}

\begin{table*}[h]
\centering
\small
\caption{Finite-sample operating characteristics of the LWR estimators under the null.}
\label{tab:five_supp_lwr_null}
\setlength{\tabcolsep}{3pt}
\renewcommand{\arraystretch}{1.08}
\begin{tabular}{@{}llrrrrrcc@{}}
\toprule
\multirow[c]{2}{*}{$N$} & \multirow[c]{2}{*}{Working set} & \multicolumn{1}{c}{Point Est.} & \multicolumn{4}{c}{Precision \& efficiency} & \multicolumn{2}{c}{Inference, \%} \\
\cmidrule(lr){3-3} \cmidrule(lr){4-7} \cmidrule(lr){8-9}
 & &   \shortstack[c]{Bias\\($\times10^{-3}$)} & ESD & ASE & ASE/ESD & PVR, \% & \shortstack[c]{CProb\\(95\% MC CI)} & \shortstack[c]{Type I error\\(95\% MC CI)} \\
\midrule
\multirow[c]{8}{*}{200} & Unadjusted & -2.93 & 0.1713 & 0.1699 & 0.992 & -- & 95.12 (94.68, 95.53) & 4.88 (4.47, 5.32) \\
 & Shared & -1.18 & 0.1501 & 0.1475 & 0.983 & 23.2 & 94.78 (94.33, 95.20) & 5.22 (4.80, 5.67) \\
 & $D_1$-complete & -1.57 & 0.1460 & 0.1418 & 0.972 & 27.4 & 94.41 (93.94, 94.84) & 5.59 (5.16, 6.06) \\
 & $D_2$-complete & -1.17 & 0.1476 & 0.1443 & 0.977 & 25.8 & 94.54 (94.08, 94.97) & 5.46 (5.03, 5.92) \\
 & $D_3$-complete & -1.09 & 0.1486 & 0.1455 & 0.979 & 24.7 & 94.69 (94.23, 95.11) & 5.31 (4.89, 5.77) \\
 & Full & -1.52 & 0.1454 & 0.1406 & 0.967 & 28.0 & 94.33 (93.86, 94.77) & 5.67 (5.23, 6.14) \\
 & Full + 5 & -1.55 & 0.1470 & 0.1388 & 0.944 & 26.4 & 93.85 (93.36, 94.30) & 6.15 (5.70, 6.64) \\
 & Full + 10 & -1.77 & 0.1484 & 0.1369 & 0.922 & 24.9 & 93.31 (92.80, 93.78) & 6.69 (6.22, 7.20) \\
\midrule
\multirow[c]{8}{*}{600} & Unadjusted & -1.12 & 0.0973 & 0.0977 & 1.004 & -- & 95.17 (94.73, 95.57) & 4.83 (4.43, 5.27) \\
 & Shared & -0.73 & 0.0848 & 0.0851 & 1.004 & 24.0 & 95.14 (94.70, 95.54) & 4.86 (4.46, 5.30) \\
 & $D_1$-complete & -0.73 & 0.0821 & 0.0821 & 1.000 & 28.7 & 95.15 (94.71, 95.55) & 4.85 (4.45, 5.29) \\
 & $D_2$-complete & -0.57 & 0.0831 & 0.0834 & 1.003 & 27.0 & 95.23 (94.79, 95.63) & 4.77 (4.37, 5.21) \\
 & $D_3$-complete & -0.73 & 0.0840 & 0.0841 & 1.001 & 25.5 & 94.98 (94.53, 95.39) & 5.02 (4.61, 5.47) \\
 & Full & -0.71 & 0.0817 & 0.0815 & 0.998 & 29.5 & 94.98 (94.53, 95.39) & 5.02 (4.61, 5.47) \\
 & Full + 5 & -0.52 & 0.0819 & 0.0812 & 0.991 & 29.1 & 94.74 (94.29, 95.16) & 5.26 (4.84, 5.71) \\
 & Full + 10 & -0.53 & 0.0822 & 0.0808 & 0.983 & 28.5 & 94.57 (94.11, 95.00) & 5.43 (5.00, 5.89) \\
\midrule
\multirow[c]{8}{*}{800} & Unadjusted & 0.69 & 0.0846 & 0.0845 & 1.000 & -- & 94.99 (94.54, 95.40) & 5.01 (4.60, 5.46) \\
 & Shared & -0.44 & 0.0740 & 0.0737 & 0.995 & 23.4 & 94.82 (94.37, 95.24) & 5.18 (4.76, 5.63) \\
 & $D_1$-complete & -0.66 & 0.0715 & 0.0711 & 0.995 & 28.5 & 95.03 (94.59, 95.44) & 4.97 (4.56, 5.41) \\
 & $D_2$-complete & -0.67 & 0.0726 & 0.0722 & 0.995 & 26.4 & 95.09 (94.65, 95.50) & 4.91 (4.50, 5.35) \\
 & $D_3$-complete & -0.39 & 0.0732 & 0.0728 & 0.995 & 25.1 & 94.75 (94.30, 95.17) & 5.25 (4.83, 5.70) \\
 & Full & -0.59 & 0.0710 & 0.0706 & 0.994 & 29.4 & 95.00 (94.56, 95.41) & 5.00 (4.59, 5.44) \\
 & Full + 5 & -0.62 & 0.0712 & 0.0704 & 0.988 & 29.0 & 94.85 (94.40, 95.27) & 5.15 (4.73, 5.60) \\
 & Full + 10 & -0.54 & 0.0714 & 0.0702 & 0.982 & 28.6 & 94.73 (94.27, 95.15) & 5.27 (4.85, 5.73) \\
\midrule
\multirow[c]{8}{*}{1200} & Unadjusted & -0.30 & 0.0683 & 0.0690 & 1.009 & -- & 95.36 (94.93, 95.76) & 4.64 (4.24, 5.07) \\
 & Shared & -0.38 & 0.0598 & 0.0601 & 1.006 & 23.5 & 95.21 (94.77, 95.61) & 4.79 (4.39, 5.23) \\
 & $D_1$-complete & -0.55 & 0.0575 & 0.0581 & 1.010 & 29.2 & 95.12 (94.68, 95.53) & 4.88 (4.47, 5.32) \\
 & $D_2$-complete & -0.44 & 0.0584 & 0.0589 & 1.009 & 26.8 & 95.14 (94.70, 95.54) & 4.86 (4.46, 5.30) \\
 & $D_3$-complete & -0.49 & 0.0589 & 0.0594 & 1.009 & 25.7 & 95.16 (94.72, 95.56) & 4.84 (4.44, 5.28) \\
 & Full & -0.62 & 0.0570 & 0.0577 & 1.011 & 30.3 & 95.21 (94.77, 95.61) & 4.79 (4.39, 5.23) \\
 & Full + 5 & -0.64 & 0.0571 & 0.0576 & 1.007 & 30.1 & 95.12 (94.68, 95.53) & 4.88 (4.47, 5.32) \\
 & Full + 10 & -0.65 & 0.0572 & 0.0574 & 1.004 & 29.8 & 95.14 (94.70, 95.54) & 4.86 (4.46, 5.30) \\
\bottomrule
\end{tabular}
\end{table*}

\begin{table*}[h]
\centering
\small
\caption{Finite-sample operating characteristics of the LWR estimators under the alternative.}
\label{tab:five_supp_lwr_alternative}
\setlength{\tabcolsep}{3pt}
\renewcommand{\arraystretch}{1.08}
\begin{tabular}{@{}llrrrrrcc@{}}
\toprule
\multirow[c]{2}{*}{$N$} & \multirow[c]{2}{*}{Working set} & \multicolumn{1}{c}{Point Est.} & \multicolumn{4}{c}{Precision \& efficiency} & \multicolumn{2}{c}{Inference, \%} \\
\cmidrule(lr){3-3} \cmidrule(lr){4-7} \cmidrule(lr){8-9}
 & & \shortstack[c]{Bias\\($\times10^{-3}$)}& ESD & ASE & ASE/ESD & PVR, \% & \shortstack[c]{CProb\\(95\% MC CI)} & \shortstack[c]{Power\\(95\% MC CI)} \\
\midrule
\multirow[c]{8}{*}{200} & Unadjusted & 3.73 & 0.1716 & 0.1707 & 0.995 & -- & 95.13 (94.69, 95.53) & 21.48 (20.69, 22.30) \\
 & Shared & 2.81 & 0.1497 & 0.1480 & 0.989 & 23.9 & 94.85 (94.40, 95.27) & 27.64 (26.77, 28.52) \\
 & $D_1$-complete & 2.90 & 0.1456 & 0.1425 & 0.978 & 28.0 & 94.32 (93.85, 94.76) & 29.28 (28.40, 30.18) \\
 & $D_2$-complete & 3.06 & 0.1472 & 0.1447 & 0.983 & 26.4 & 94.49 (94.03, 94.92) & 28.73 (27.85, 29.62) \\
 & $D_3$-complete & 3.17 & 0.1483 & 0.1459 & 0.983 & 25.3 & 94.67 (94.21, 95.09) & 28.25 (27.38, 29.14) \\
 & Full & 3.22 & 0.1450 & 0.1412 & 0.974 & 28.6 & 94.19 (93.71, 94.63) & 29.80 (28.91, 30.70) \\
 & Full + 5 & 3.44 & 0.1467 & 0.1394 & 0.950 & 26.9 & 93.50 (93.00, 93.97) & 30.85 (29.95, 31.76) \\
 & Full + 10 & 3.76 & 0.1482 & 0.1375 & 0.928 & 25.4 & 93.06 (92.55, 93.54) & 32.04 (31.13, 32.96) \\
\midrule
\multirow[c]{8}{*}{600} & Unadjusted & -0.17 & 0.0981 & 0.0981 & 1.000 & -- & 95.15 (94.71, 95.55) & 51.72 (50.74, 52.70) \\
 & Shared & -0.32 & 0.0860 & 0.0853 & 0.992 & 23.3 & 94.73 (94.27, 95.15) & 63.88 (62.93, 64.82) \\
 & $D_1$-complete & -0.23 & 0.0835 & 0.0824 & 0.987 & 27.7 & 94.66 (94.20, 95.08) & 66.73 (65.80, 67.65) \\
 & $D_2$-complete & -0.31 & 0.0844 & 0.0835 & 0.990 & 26.1 & 94.58 (94.12, 95.01) & 65.50 (64.56, 66.43) \\
 & $D_3$-complete & -0.40 & 0.0850 & 0.0842 & 0.991 & 25.0 & 94.76 (94.31, 95.18) & 64.87 (63.93, 65.80) \\
 & Full & -0.26 & 0.0830 & 0.0818 & 0.985 & 28.5 & 94.55 (94.09, 94.98) & 67.38 (66.45, 68.29) \\
 & Full + 5 & -0.28 & 0.0833 & 0.0814 & 0.978 & 27.9 & 94.40 (93.93, 94.83) & 67.80 (66.88, 68.71) \\
 & Full + 10 & -0.15 & 0.0836 & 0.0811 & 0.970 & 27.4 & 94.25 (93.78, 94.69) & 68.13 (67.21, 69.04) \\
\midrule
\multirow[c]{8}{*}{800} & Unadjusted & 1.56 & 0.0847 & 0.0849 & 1.003 & -- & 95.17 (94.73, 95.57) & 64.87 (63.93, 65.80) \\
 & Shared & 1.83 & 0.0740 & 0.0739 & 0.999 & 23.7 & 94.99 (94.54, 95.40) & 77.13 (76.30, 77.94) \\
 & $D_1$-complete & 1.57 & 0.0717 & 0.0714 & 0.996 & 28.4 & 94.78 (94.33, 95.20) & 79.74 (78.94, 80.52) \\
 & $D_2$-complete & 1.70 & 0.0726 & 0.0724 & 0.997 & 26.6 & 94.83 (94.38, 95.25) & 78.92 (78.11, 79.71) \\
 & $D_3$-complete & 1.77 & 0.0729 & 0.0729 & 1.001 & 26.0 & 95.14 (94.70, 95.54) & 78.39 (77.57, 79.19) \\
 & Full & 1.56 & 0.0711 & 0.0709 & 0.997 & 29.6 & 94.84 (94.39, 95.26) & 80.57 (79.78, 81.33) \\
 & Full + 5 & 1.66 & 0.0714 & 0.0707 & 0.990 & 29.0 & 94.70 (94.24, 95.12) & 80.56 (79.77, 81.32) \\
 & Full + 10 & 1.63 & 0.0716 & 0.0704 & 0.983 & 28.5 & 94.51 (94.05, 94.94) & 80.50 (79.71, 81.26) \\
\midrule
\multirow[c]{8}{*}{1200} & Unadjusted & 1.50 & 0.0690 & 0.0693 & 1.004 & -- & 94.76 (94.31, 95.18) & 82.48 (81.72, 83.21) \\
 & Shared & 0.94 & 0.0603 & 0.0603 & 1.000 & 23.7 & 94.96 (94.51, 95.37) & 91.18 (90.61, 91.72) \\
 & $D_1$-complete & 0.87 & 0.0584 & 0.0583 & 0.998 & 28.3 & 94.94 (94.49, 95.35) & 92.79 (92.27, 93.28) \\
 & $D_2$-complete & 0.82 & 0.0592 & 0.0591 & 0.998 & 26.4 & 94.91 (94.46, 95.32) & 92.21 (91.67, 92.72) \\
 & $D_3$-complete & 1.02 & 0.0595 & 0.0595 & 1.001 & 25.7 & 95.02 (94.58, 95.43) & 91.91 (91.36, 92.43) \\
 & Full & 0.94 & 0.0580 & 0.0579 & 0.998 & 29.4 & 95.05 (94.61, 95.46) & 93.16 (92.65, 93.64) \\
 & Full + 5 & 0.93 & 0.0581 & 0.0578 & 0.995 & 29.2 & 94.97 (94.52, 95.38) & 93.16 (92.65, 93.64) \\
 & Full + 10 & 1.00 & 0.0582 & 0.0576 & 0.991 & 29.0 & 94.89 (94.44, 95.30) & 93.32 (92.81, 93.79) \\
\bottomrule
\end{tabular}
\end{table*}

\begin{table*}[h]
\centering
\small
\caption{Finite-sample operating characteristics of the NB estimators under the null.}
\label{tab:five_supp_nb_null}
\setlength{\tabcolsep}{3pt}
\renewcommand{\arraystretch}{1.08}
\begin{tabular}{@{}llrrrrrcc@{}}
\toprule
\multirow[c]{2}{*}{$N$} & \multirow[c]{2}{*}{Working set} & \multicolumn{1}{c}{Point Est.} & \multicolumn{4}{c}{Precision \& efficiency} & \multicolumn{2}{c}{Inference, \%} \\
\cmidrule(lr){3-3} \cmidrule(lr){4-7} \cmidrule(lr){8-9}
 & & \shortstack[c]{Bias\\($\times10^{-3}$)}& ESD & ASE & ASE/ESD & PVR, \% & \shortstack[c]{CProb\\(95\% MC CI)} & \shortstack[c]{Type I error\\(95\% MC CI)} \\
\midrule
\multirow[c]{8}{*}{200} & Unadjusted & -1.38 & 0.0811 & 0.0804 & 0.992 & -- & 94.53 (94.07, 94.96) & 5.47 (5.04, 5.93) \\
 & Shared & -0.55 & 0.0711 & 0.0699 & 0.983 & 23.0 & 94.56 (94.10, 94.99) & 5.44 (5.01, 5.90) \\
 & $D_1$-complete & -0.75 & 0.0692 & 0.0672 & 0.972 & 27.2 & 94.12 (93.64, 94.56) & 5.88 (5.44, 6.36) \\
 & $D_2$-complete & -0.55 & 0.0699 & 0.0684 & 0.977 & 25.6 & 94.22 (93.75, 94.66) & 5.78 (5.34, 6.25) \\
 & $D_3$-complete & -0.51 & 0.0704 & 0.0689 & 0.979 & 24.5 & 94.32 (93.85, 94.76) & 5.68 (5.24, 6.15) \\
 & Full & -0.72 & 0.0689 & 0.0666 & 0.967 & 27.8 & 94.12 (93.64, 94.56) & 5.88 (5.44, 6.36) \\
 & Full + 5 & -0.74 & 0.0697 & 0.0658 & 0.944 & 26.2 & 93.53 (93.03, 94.00) & 6.47 (6.00, 6.97) \\
 & Full + 10 & -0.84 & 0.0703 & 0.0649 & 0.923 & 24.7 & 93.03 (92.51, 93.51) & 6.97 (6.49, 7.49) \\
\midrule
\multirow[c]{8}{*}{600} & Unadjusted & -0.53 & 0.0462 & 0.0464 & 1.004 & -- & 95.01 (94.57, 95.42) & 4.99 (4.58, 5.43) \\
 & Shared & -0.35 & 0.0403 & 0.0405 & 1.004 & 24.0 & 95.09 (94.65, 95.50) & 4.91 (4.50, 5.35) \\
 & $D_1$-complete & -0.35 & 0.0391 & 0.0391 & 1.000 & 28.6 & 95.12 (94.68, 95.53) & 4.88 (4.47, 5.32) \\
 & $D_2$-complete & -0.27 & 0.0395 & 0.0396 & 1.003 & 26.9 & 95.13 (94.69, 95.53) & 4.87 (4.47, 5.31) \\
 & $D_3$-complete & -0.35 & 0.0399 & 0.0400 & 1.001 & 25.4 & 94.92 (94.47, 95.33) & 5.08 (4.67, 5.53) \\
 & Full & -0.34 & 0.0389 & 0.0388 & 0.998 & 29.4 & 94.88 (94.43, 95.30) & 5.12 (4.70, 5.57) \\
 & Full + 5 & -0.25 & 0.0390 & 0.0386 & 0.991 & 29.0 & 94.68 (94.22, 95.10) & 5.32 (4.90, 5.78) \\
 & Full + 10 & -0.25 & 0.0391 & 0.0385 & 0.983 & 28.5 & 94.50 (94.04, 94.93) & 5.50 (5.07, 5.96) \\
\midrule
\multirow[c]{8}{*}{800} & Unadjusted & 0.33 & 0.0402 & 0.0402 & 1.000 & -- & 94.92 (94.47, 95.33) & 5.08 (4.67, 5.53) \\
 & Shared & -0.21 & 0.0352 & 0.0351 & 0.995 & 23.3 & 94.79 (94.34, 95.21) & 5.21 (4.79, 5.66) \\
 & $D_1$-complete & -0.31 & 0.0340 & 0.0338 & 0.995 & 28.5 & 94.98 (94.53, 95.39) & 5.02 (4.61, 5.47) \\
 & $D_2$-complete & -0.32 & 0.0345 & 0.0344 & 0.995 & 26.3 & 95.02 (94.58, 95.43) & 4.98 (4.57, 5.42) \\
 & $D_3$-complete & -0.18 & 0.0348 & 0.0346 & 0.995 & 25.1 & 94.66 (94.20, 95.08) & 5.34 (4.92, 5.80) \\
 & Full & -0.28 & 0.0338 & 0.0336 & 0.994 & 29.4 & 94.95 (94.50, 95.36) & 5.05 (4.64, 5.50) \\
 & Full + 5 & -0.29 & 0.0339 & 0.0335 & 0.988 & 29.0 & 94.78 (94.33, 95.20) & 5.22 (4.80, 5.67) \\
 & Full + 10 & -0.25 & 0.0340 & 0.0334 & 0.982 & 28.6 & 94.65 (94.19, 95.07) & 5.35 (4.93, 5.81) \\
\midrule
\multirow[c]{8}{*}{1200} & Unadjusted & -0.14 & 0.0325 & 0.0328 & 1.009 & -- & 95.26 (94.83, 95.66) & 4.74 (4.34, 5.17) \\
 & Shared & -0.18 & 0.0285 & 0.0286 & 1.006 & 23.5 & 95.20 (94.76, 95.60) & 4.80 (4.40, 5.24) \\
 & $D_1$-complete & -0.26 & 0.0274 & 0.0276 & 1.010 & 29.1 & 95.10 (94.66, 95.51) & 4.90 (4.49, 5.34) \\
 & $D_2$-complete & -0.21 & 0.0278 & 0.0281 & 1.009 & 26.8 & 95.09 (94.65, 95.50) & 4.91 (4.50, 5.35) \\
 & $D_3$-complete & -0.23 & 0.0280 & 0.0283 & 1.009 & 25.7 & 95.11 (94.67, 95.52) & 4.89 (4.48, 5.33) \\
 & Full & -0.30 & 0.0272 & 0.0275 & 1.011 & 30.3 & 95.15 (94.71, 95.55) & 4.85 (4.45, 5.29) \\
 & Full + 5 & -0.31 & 0.0272 & 0.0274 & 1.007 & 30.0 & 95.09 (94.65, 95.50) & 4.91 (4.50, 5.35) \\
 & Full + 10 & -0.31 & 0.0272 & 0.0273 & 1.004 & 29.8 & 95.07 (94.63, 95.48) & 4.93 (4.52, 5.37) \\
\bottomrule
\end{tabular}
\end{table*}

\begin{table*}[h]
\centering
\small
\caption{Finite-sample operating characteristics of the NB estimators under the alternative.}
\label{tab:five_supp_nb_alternative}
\setlength{\tabcolsep}{3pt}
\renewcommand{\arraystretch}{1.08}
\begin{tabular}{@{}llrrrrrcc@{}}
\toprule
\multirow[c]{2}{*}{$N$} & \multirow[c]{2}{*}{Working set} & \multicolumn{1}{c}{Point Est.} & \multicolumn{4}{c}{Precision \& efficiency} & \multicolumn{2}{c}{Inference, \%} \\
\cmidrule(lr){3-3} \cmidrule(lr){4-7} \cmidrule(lr){8-9}
 & & \shortstack[c]{Bias\\($\times10^{-3}$)}& ESD & ASE & ASE/ESD & PVR, \% & \shortstack[c]{CProb\\(95\% MC CI)} & \shortstack[c]{Power\\(95\% MC CI)} \\
\midrule
\multirow[c]{8}{*}{200} & Unadjusted & 1.10 & 0.0803 & 0.0799 & 0.995 & -- & 94.66 (94.20, 95.08) & 22.48 (21.67, 23.31) \\
 & Shared & 0.80 & 0.0701 & 0.0693 & 0.989 & 23.7 & 94.56 (94.10, 94.99) & 28.32 (27.45, 29.21) \\
 & $D_1$-complete & 0.86 & 0.0682 & 0.0668 & 0.978 & 27.7 & 94.12 (93.64, 94.56) & 30.10 (29.21, 31.01) \\
 & $D_2$-complete & 0.93 & 0.0690 & 0.0678 & 0.983 & 26.1 & 94.19 (93.71, 94.63) & 29.47 (28.58, 30.37) \\
 & $D_3$-complete & 0.96 & 0.0695 & 0.0683 & 0.984 & 25.1 & 94.23 (93.76, 94.67) & 28.96 (28.08, 29.86) \\
 & Full & 1.01 & 0.0679 & 0.0662 & 0.974 & 28.4 & 93.94 (93.46, 94.39) & 30.51 (29.62, 31.42) \\
 & Full + 5 & 1.10 & 0.0687 & 0.0653 & 0.950 & 26.7 & 93.21 (92.70, 93.69) & 31.56 (30.66, 32.48) \\
 & Full + 10 & 1.25 & 0.0694 & 0.0644 & 0.928 & 25.2 & 92.65 (92.12, 93.15) & 32.81 (31.90, 33.74) \\
\midrule
\multirow[c]{8}{*}{600} & Unadjusted & -0.29 & 0.0461 & 0.0461 & 1.000 & -- & 95.02 (94.58, 95.43) & 52.09 (51.11, 53.07) \\
 & Shared & -0.31 & 0.0404 & 0.0401 & 0.992 & 23.2 & 94.62 (94.16, 95.05) & 64.19 (63.25, 65.12) \\
 & $D_1$-complete & -0.26 & 0.0393 & 0.0388 & 0.987 & 27.6 & 94.60 (94.14, 95.03) & 67.09 (66.16, 68.00) \\
 & $D_2$-complete & -0.31 & 0.0397 & 0.0393 & 0.989 & 26.0 & 94.52 (94.06, 94.95) & 65.92 (64.99, 66.84) \\
 & $D_3$-complete & -0.35 & 0.0400 & 0.0396 & 0.991 & 25.0 & 94.69 (94.23, 95.11) & 65.25 (64.31, 66.18) \\
 & Full & -0.28 & 0.0390 & 0.0385 & 0.985 & 28.4 & 94.48 (94.02, 94.91) & 67.65 (66.73, 68.56) \\
 & Full + 5 & -0.29 & 0.0392 & 0.0383 & 0.977 & 27.9 & 94.39 (93.92, 94.82) & 68.09 (67.17, 69.00) \\
 & Full + 10 & -0.23 & 0.0394 & 0.0381 & 0.969 & 27.3 & 94.04 (93.56, 94.49) & 68.38 (67.46, 69.28) \\
\midrule
\multirow[c]{8}{*}{800} & Unadjusted & 0.58 & 0.0398 & 0.0400 & 1.003 & -- & 95.11 (94.67, 95.52) & 65.18 (64.24, 66.11) \\
 & Shared & 0.74 & 0.0348 & 0.0348 & 0.999 & 23.7 & 94.89 (94.44, 95.30) & 77.40 (76.57, 78.21) \\
 & $D_1$-complete & 0.63 & 0.0337 & 0.0336 & 0.996 & 28.4 & 94.70 (94.24, 95.12) & 79.95 (79.15, 80.72) \\
 & $D_2$-complete & 0.69 & 0.0341 & 0.0340 & 0.997 & 26.6 & 94.74 (94.29, 95.16) & 79.11 (78.30, 79.90) \\
 & $D_3$-complete & 0.72 & 0.0343 & 0.0343 & 1.001 & 25.9 & 95.11 (94.67, 95.52) & 78.60 (77.79, 79.39) \\
 & Full & 0.62 & 0.0334 & 0.0334 & 0.997 & 29.5 & 94.77 (94.32, 95.19) & 80.73 (79.95, 81.49) \\
 & Full + 5 & 0.67 & 0.0336 & 0.0332 & 0.990 & 28.9 & 94.60 (94.14, 95.03) & 80.68 (79.89, 81.44) \\
 & Full + 10 & 0.66 & 0.0337 & 0.0331 & 0.984 & 28.5 & 94.40 (93.93, 94.83) & 80.74 (79.96, 81.50) \\
\midrule
\multirow[c]{8}{*}{1200} & Unadjusted & 0.61 & 0.0325 & 0.0326 & 1.004 & -- & 94.70 (94.24, 95.12) & 82.59 (81.83, 83.32) \\
 & Shared & 0.36 & 0.0284 & 0.0284 & 1.000 & 23.7 & 94.91 (94.46, 95.32) & 91.22 (90.65, 91.76) \\
 & $D_1$-complete & 0.34 & 0.0275 & 0.0274 & 0.998 & 28.3 & 94.81 (94.36, 95.23) & 92.82 (92.30, 93.31) \\
 & $D_2$-complete & 0.31 & 0.0279 & 0.0278 & 0.998 & 26.4 & 94.81 (94.36, 95.23) & 92.25 (91.71, 92.76) \\
 & $D_3$-complete & 0.40 & 0.0280 & 0.0280 & 1.001 & 25.7 & 94.96 (94.51, 95.37) & 92.00 (91.45, 92.52) \\
 & Full & 0.37 & 0.0273 & 0.0272 & 0.998 & 29.3 & 95.04 (94.60, 95.45) & 93.20 (92.69, 93.68) \\
 & Full + 5 & 0.36 & 0.0273 & 0.0272 & 0.995 & 29.2 & 94.89 (94.44, 95.30) & 93.25 (92.74, 93.73) \\
 & Full + 10 & 0.39 & 0.0274 & 0.0271 & 0.991 & 29.0 & 94.81 (94.36, 95.23) & 93.34 (92.83, 93.81) \\
\bottomrule
\end{tabular}
\end{table*}

\begin{table*}[h]
\centering
\small
\caption{Finite-sample operating characteristics of the LWO estimators under the null.}
\label{tab:five_supp_lwo_null}
\setlength{\tabcolsep}{3pt}
\renewcommand{\arraystretch}{1.08}
\begin{tabular}{@{}llrrrrrcc@{}}
\toprule
\multirow[c]{2}{*}{$N$} & \multirow[c]{2}{*}{Working set} & \multicolumn{1}{c}{Point Est.} & \multicolumn{4}{c}{Precision \& efficiency} & \multicolumn{2}{c}{Inference, \%} \\
\cmidrule(lr){3-3} \cmidrule(lr){4-7} \cmidrule(lr){8-9}
 & & \shortstack[c]{Bias\\($\times10^{-3}$)}& ESD & ASE & ASE/ESD & PVR, \% & \shortstack[c]{CProb\\(95\% MC CI)} & \shortstack[c]{Type I error\\(95\% MC CI)} \\
\midrule
\multirow[c]{8}{*}{200} & Unadjusted & -2.77 & 0.1632 & 0.1618 & 0.992 & -- & 95.11 (94.67, 95.52) & 4.89 (4.48, 5.33) \\
 & Shared & -1.11 & 0.1430 & 0.1405 & 0.983 & 23.2 & 94.79 (94.34, 95.21) & 5.21 (4.79, 5.66) \\
 & $D_1$-complete & -1.49 & 0.1390 & 0.1351 & 0.972 & 27.4 & 94.41 (93.94, 94.84) & 5.59 (5.16, 6.06) \\
 & $D_2$-complete & -1.10 & 0.1406 & 0.1374 & 0.978 & 25.8 & 94.53 (94.07, 94.96) & 5.47 (5.04, 5.93) \\
 & $D_3$-complete & -1.03 & 0.1415 & 0.1386 & 0.979 & 24.8 & 94.67 (94.21, 95.09) & 5.33 (4.91, 5.79) \\
 & Full & -1.45 & 0.1384 & 0.1339 & 0.967 & 28.0 & 94.33 (93.86, 94.77) & 5.67 (5.23, 6.14) \\
 & Full + 5 & -1.47 & 0.1400 & 0.1322 & 0.944 & 26.4 & 93.85 (93.36, 94.30) & 6.15 (5.70, 6.64) \\
 & Full + 10 & -1.68 & 0.1413 & 0.1304 & 0.923 & 25.0 & 93.31 (92.80, 93.78) & 6.69 (6.22, 7.20) \\
\midrule
\multirow[c]{8}{*}{600} & Unadjusted & -1.07 & 0.0927 & 0.0930 & 1.004 & -- & 95.17 (94.73, 95.57) & 4.83 (4.43, 5.27) \\
 & Shared & -0.70 & 0.0808 & 0.0811 & 1.004 & 24.1 & 95.14 (94.70, 95.54) & 4.86 (4.46, 5.30) \\
 & $D_1$-complete & -0.70 & 0.0783 & 0.0782 & 1.000 & 28.7 & 95.15 (94.71, 95.55) & 4.85 (4.45, 5.29) \\
 & $D_2$-complete & -0.55 & 0.0792 & 0.0794 & 1.003 & 27.0 & 95.22 (94.78, 95.62) & 4.78 (4.38, 5.22) \\
 & $D_3$-complete & -0.70 & 0.0800 & 0.0801 & 1.001 & 25.5 & 94.98 (94.53, 95.39) & 5.02 (4.61, 5.47) \\
 & Full & -0.68 & 0.0778 & 0.0777 & 0.998 & 29.5 & 94.97 (94.52, 95.38) & 5.03 (4.62, 5.48) \\
 & Full + 5 & -0.50 & 0.0780 & 0.0774 & 0.991 & 29.1 & 94.72 (94.26, 95.14) & 5.28 (4.86, 5.74) \\
 & Full + 10 & -0.51 & 0.0783 & 0.0770 & 0.983 & 28.6 & 94.57 (94.11, 95.00) & 5.43 (5.00, 5.89) \\
\midrule
\multirow[c]{8}{*}{800} & Unadjusted & 0.66 & 0.0806 & 0.0805 & 1.000 & -- & 94.99 (94.54, 95.40) & 5.01 (4.60, 5.46) \\
 & Shared & -0.42 & 0.0705 & 0.0702 & 0.995 & 23.4 & 94.82 (94.37, 95.24) & 5.18 (4.76, 5.63) \\
 & $D_1$-complete & -0.62 & 0.0681 & 0.0678 & 0.995 & 28.5 & 95.03 (94.59, 95.44) & 4.97 (4.56, 5.41) \\
 & $D_2$-complete & -0.63 & 0.0691 & 0.0688 & 0.995 & 26.4 & 95.09 (94.65, 95.50) & 4.91 (4.50, 5.35) \\
 & $D_3$-complete & -0.37 & 0.0697 & 0.0693 & 0.995 & 25.1 & 94.74 (94.29, 95.16) & 5.26 (4.84, 5.71) \\
 & Full & -0.56 & 0.0677 & 0.0673 & 0.994 & 29.4 & 95.00 (94.56, 95.41) & 5.00 (4.59, 5.44) \\
 & Full + 5 & -0.59 & 0.0679 & 0.0671 & 0.988 & 29.0 & 94.84 (94.39, 95.26) & 5.16 (4.74, 5.61) \\
 & Full + 10 & -0.51 & 0.0681 & 0.0669 & 0.982 & 28.6 & 94.73 (94.27, 95.15) & 5.27 (4.85, 5.73) \\
\midrule
\multirow[c]{8}{*}{1200} & Unadjusted & -0.29 & 0.0651 & 0.0657 & 1.009 & -- & 95.36 (94.93, 95.76) & 4.64 (4.24, 5.07) \\
 & Shared & -0.36 & 0.0569 & 0.0573 & 1.006 & 23.5 & 95.21 (94.77, 95.61) & 4.79 (4.39, 5.23) \\
 & $D_1$-complete & -0.53 & 0.0548 & 0.0553 & 1.010 & 29.2 & 95.12 (94.68, 95.53) & 4.88 (4.47, 5.32) \\
 & $D_2$-complete & -0.42 & 0.0557 & 0.0562 & 1.009 & 26.9 & 95.14 (94.70, 95.54) & 4.86 (4.46, 5.30) \\
 & $D_3$-complete & -0.47 & 0.0561 & 0.0566 & 1.009 & 25.7 & 95.16 (94.72, 95.56) & 4.84 (4.44, 5.28) \\
 & Full & -0.60 & 0.0544 & 0.0550 & 1.011 & 30.3 & 95.21 (94.77, 95.61) & 4.79 (4.39, 5.23) \\
 & Full + 5 & -0.61 & 0.0544 & 0.0549 & 1.007 & 30.1 & 95.12 (94.68, 95.53) & 4.88 (4.47, 5.32) \\
 & Full + 10 & -0.62 & 0.0545 & 0.0547 & 1.004 & 29.8 & 95.12 (94.68, 95.53) & 4.88 (4.47, 5.32) \\
\bottomrule
\end{tabular}
\end{table*}

\begin{table*}[h]
\centering
\small
\caption{Finite-sample operating characteristics of the LWO estimators under the alternative.}
\label{tab:five_supp_lwo_alternative}
\setlength{\tabcolsep}{3pt}
\renewcommand{\arraystretch}{1.08}
\begin{tabular}{@{}llrrrrrcc@{}}
\toprule
\multirow[c]{2}{*}{$N$} & \multirow[c]{2}{*}{Working set} & \multicolumn{1}{c}{Point Est.} & \multicolumn{4}{c}{Precision \& efficiency} & \multicolumn{2}{c}{Inference, \%} \\
\cmidrule(lr){3-3} \cmidrule(lr){4-7} \cmidrule(lr){8-9}
 & & \shortstack[c]{Bias\\($\times10^{-3}$)}& ESD & ASE & ASE/ESD & PVR, \% & \shortstack[c]{CProb\\(95\% MC CI)} & \shortstack[c]{Power\\(95\% MC CI)} \\
\midrule
\multirow[c]{8}{*}{200} & Unadjusted & 3.48 & 0.1631 & 0.1623 & 0.995 & -- & 95.11 (94.67, 95.52) & 21.54 (20.75, 22.36) \\
 & Shared & 2.57 & 0.1422 & 0.1407 & 0.989 & 24.0 & 94.82 (94.37, 95.24) & 27.64 (26.77, 28.52) \\
 & $D_1$-complete & 2.65 & 0.1384 & 0.1354 & 0.978 & 28.0 & 94.32 (93.85, 94.76) & 29.30 (28.42, 30.20) \\
 & $D_2$-complete & 2.80 & 0.1399 & 0.1375 & 0.983 & 26.4 & 94.47 (94.00, 94.90) & 28.78 (27.90, 29.68) \\
 & $D_3$-complete & 2.88 & 0.1409 & 0.1386 & 0.984 & 25.4 & 94.63 (94.17, 95.05) & 28.28 (27.41, 29.17) \\
 & Full & 2.94 & 0.1377 & 0.1341 & 0.974 & 28.7 & 94.18 (93.70, 94.62) & 29.81 (28.92, 30.71) \\
 & Full + 5 & 3.15 & 0.1394 & 0.1324 & 0.950 & 26.9 & 93.47 (92.97, 93.94) & 30.93 (30.03, 31.84) \\
 & Full + 10 & 3.46 & 0.1408 & 0.1307 & 0.928 & 25.4 & 93.08 (92.57, 93.56) & 32.06 (31.15, 32.98) \\
\midrule
\multirow[c]{8}{*}{600} & Unadjusted & -0.17 & 0.0933 & 0.0933 & 1.000 & -- & 95.17 (94.73, 95.57) & 51.73 (50.75, 52.71) \\
 & Shared & -0.32 & 0.0817 & 0.0811 & 0.992 & 23.3 & 94.70 (94.24, 95.12) & 63.88 (62.93, 64.82) \\
 & $D_1$-complete & -0.24 & 0.0793 & 0.0783 & 0.987 & 27.7 & 94.67 (94.21, 95.09) & 66.74 (65.81, 67.66) \\
 & $D_2$-complete & -0.32 & 0.0802 & 0.0794 & 0.989 & 26.1 & 94.57 (94.11, 95.00) & 65.51 (64.57, 66.44) \\
 & $D_3$-complete & -0.41 & 0.0808 & 0.0800 & 0.991 & 25.0 & 94.75 (94.30, 95.17) & 64.87 (63.93, 65.80) \\
 & Full & -0.27 & 0.0789 & 0.0777 & 0.985 & 28.5 & 94.55 (94.09, 94.98) & 67.38 (66.45, 68.29) \\
 & Full + 5 & -0.30 & 0.0792 & 0.0774 & 0.977 & 28.0 & 94.42 (93.95, 94.85) & 67.81 (66.89, 68.72) \\
 & Full + 10 & -0.18 & 0.0795 & 0.0771 & 0.969 & 27.4 & 94.22 (93.75, 94.66) & 68.15 (67.23, 69.06) \\
\midrule
\multirow[c]{8}{*}{800} & Unadjusted & 1.49 & 0.0805 & 0.0808 & 1.003 & -- & 95.19 (94.75, 95.59) & 64.87 (63.93, 65.80) \\
 & Shared & 1.74 & 0.0703 & 0.0702 & 0.999 & 23.8 & 95.01 (94.57, 95.42) & 77.15 (76.32, 77.96) \\
 & $D_1$-complete & 1.49 & 0.0681 & 0.0679 & 0.996 & 28.5 & 94.79 (94.34, 95.21) & 79.75 (78.95, 80.53) \\
 & $D_2$-complete & 1.61 & 0.0690 & 0.0688 & 0.997 & 26.6 & 94.84 (94.39, 95.26) & 78.94 (78.13, 79.73) \\
 & $D_3$-complete & 1.67 & 0.0693 & 0.0693 & 1.001 & 26.0 & 95.16 (94.72, 95.56) & 78.41 (77.59, 79.21) \\
 & Full & 1.48 & 0.0676 & 0.0674 & 0.997 & 29.6 & 94.87 (94.42, 95.29) & 80.58 (79.79, 81.34) \\
 & Full + 5 & 1.56 & 0.0679 & 0.0672 & 0.990 & 29.0 & 94.69 (94.23, 95.11) & 80.57 (79.78, 81.33) \\
 & Full + 10 & 1.54 & 0.0681 & 0.0670 & 0.984 & 28.6 & 94.51 (94.05, 94.94) & 80.55 (79.76, 81.31) \\
\midrule
\multirow[c]{8}{*}{1200} & Unadjusted & 1.43 & 0.0656 & 0.0659 & 1.004 & -- & 94.79 (94.34, 95.21) & 82.48 (81.72, 83.21) \\
 & Shared & 0.88 & 0.0573 & 0.0573 & 1.000 & 23.8 & 94.95 (94.50, 95.36) & 91.18 (90.61, 91.72) \\
 & $D_1$-complete & 0.82 & 0.0555 & 0.0554 & 0.998 & 28.4 & 94.90 (94.45, 95.31) & 92.78 (92.26, 93.27) \\
 & $D_2$-complete & 0.77 & 0.0563 & 0.0561 & 0.998 & 26.5 & 94.87 (94.42, 95.29) & 92.21 (91.67, 92.72) \\
 & $D_3$-complete & 0.96 & 0.0565 & 0.0566 & 1.001 & 25.7 & 95.01 (94.57, 95.42) & 91.91 (91.36, 92.43) \\
 & Full & 0.88 & 0.0551 & 0.0550 & 0.998 & 29.4 & 95.11 (94.67, 95.52) & 93.16 (92.65, 93.64) \\
 & Full + 5 & 0.88 & 0.0552 & 0.0549 & 0.995 & 29.2 & 95.01 (94.57, 95.42) & 93.17 (92.66, 93.65) \\
 & Full + 10 & 0.94 & 0.0553 & 0.0548 & 0.991 & 29.0 & 94.87 (94.42, 95.29) & 93.32 (92.81, 93.79) \\
\bottomrule
\end{tabular}
\end{table*}

\begin{figure*}[h]
    \centering
    \includegraphics[width=\linewidth]{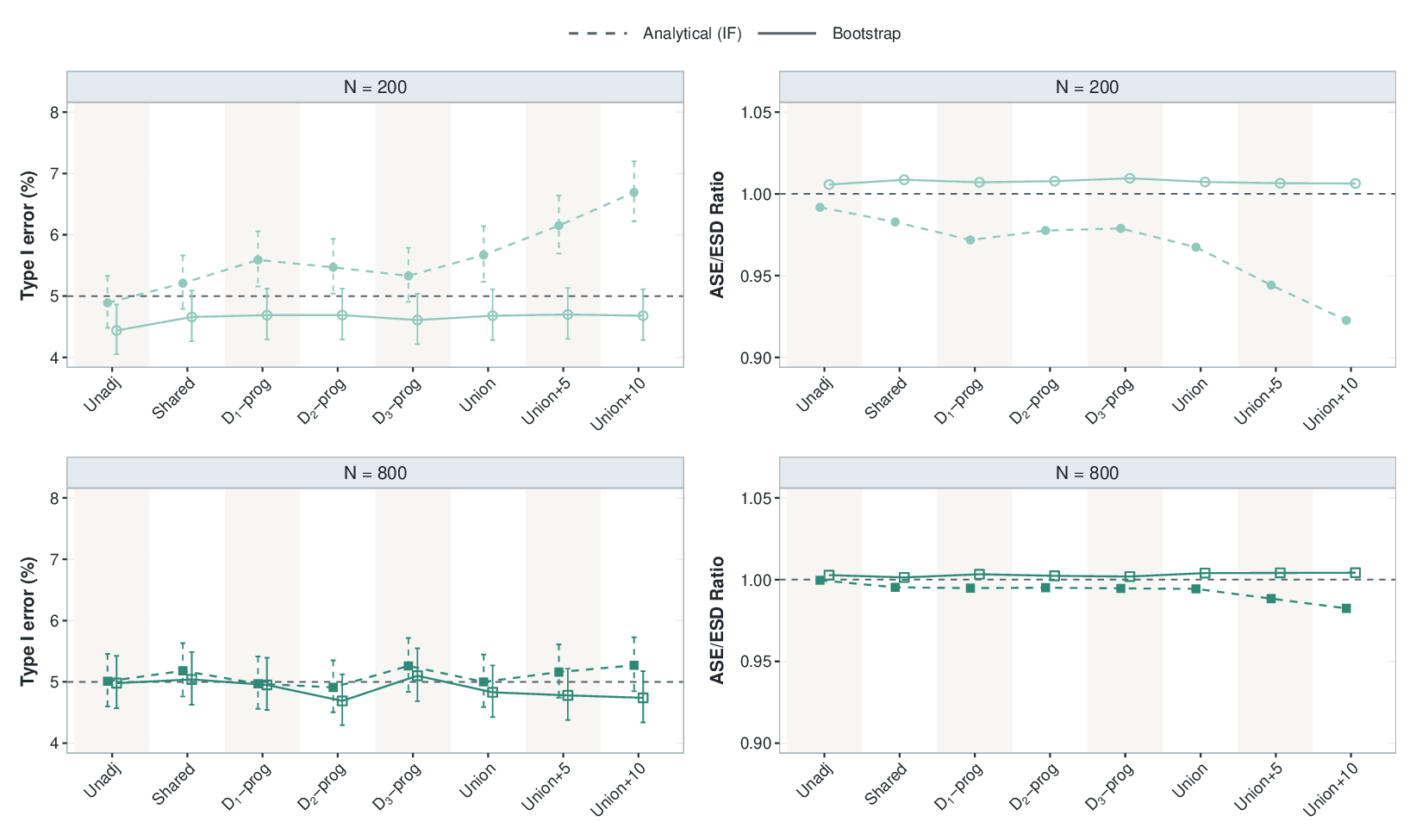}
    \caption{Comparison of analytical IF-based and bootstrap variance estimation for LWO under $H_0$ at $N=200$ (top) and $N=800$ (bottom). The left column shows Type~I error; the right shows average estimated SE divided by empirical SD (ASE/ESD). Each sample size uses the same $R=10{,}000$ original trials, with $2{,}000$ bootstrap resamples per trial. Error bars are pointwise $95\%$ Wilson Monte Carlo confidence intervals.}
    \label{fig:placeholder}
\end{figure*}

\begin{figure*}[h]
    \centering
    \includegraphics[width=\linewidth]{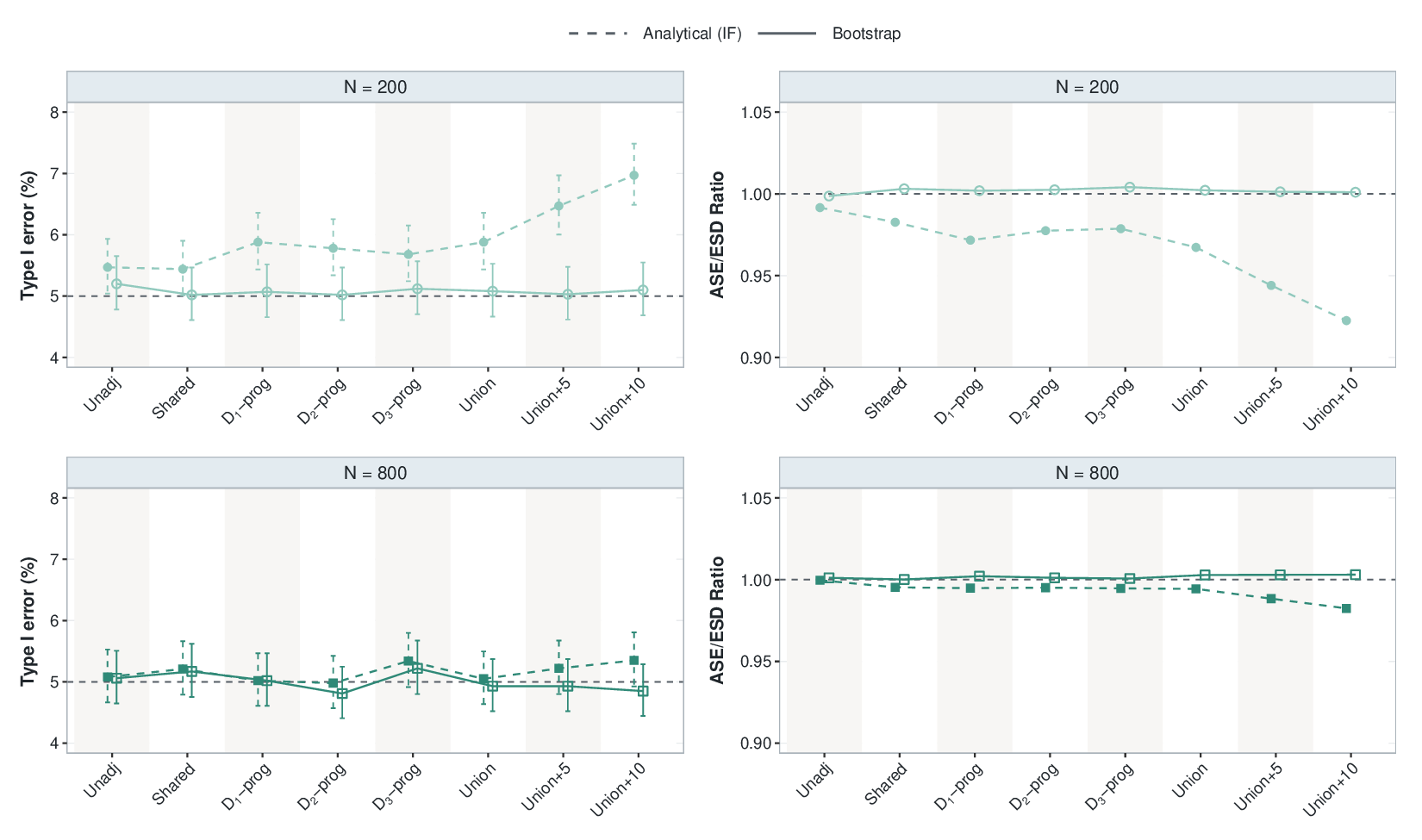}
    \caption{Comparison of analytical IF-based and bootstrap variance estimation for NB under $H_0$ at $N=200$ (top) and $N=800$ (bottom). The left column shows Type~I error; the right shows average estimated SE divided by empirical SD (ASE/ESD). Each sample size uses the same $R=10{,}000$ original trials, with $2{,}000$ bootstrap resamples per trial. Error bars are pointwise $95\%$ Wilson Monte Carlo confidence intervals.}
    \label{fig:placeholder}
\end{figure*}

\clearpage

\end{document}